\documentclass[11pt,letterpaper]{article} 

\usepackage[table]{xcolor}
\usepackage{tikz}
\usepackage{lmodern}
\usepackage{calc}
\usepackage{hyperref}
\usepackage{float}
\usepackage{multirow}
\usepackage[sort&compress,numbers]{natbib}
\usepackage[margin=1in]{geometry}
\usetikzlibrary{shapes.geometric, positioning, arrows.meta}
\usetikzlibrary{calc, fit}
\usepackage{caption}

\renewcommand{\small}{\fontsize{11}{13}\selectfont}

\usepackage{amsmath, amssymb,amsfonts,mathrsfs}
\DeclareFontFamily{U}{rsfs}{}
\DeclareFontShape{U}{rsfs}{m}{n}{<-6>rsfs5<6-8>rsfs7<8->rsfs10}{}
\usepackage{amsthm}
\usepackage{booktabs}
\usepackage{array}
\definecolor{kleinblue}{RGB}{0,47,167} 
\newtheorem{theorem}{Theorem}
\newtheorem{lemma}[theorem]{Lemma}
\newtheorem{corollary}[theorem]{Corollary}
\newtheorem{proposition}[theorem]{Proposition}
\newtheorem{definition}[theorem]{Definition}

\newcommand{\HW}{\text{HW}}

\newcommand{\B}{\mathcal{B}}
\newcommand{\F}{\mathcal{F}}
\newcommand{\G}{\mathcal{G}}
\newcommand{\Oth}{\mathcal{O}}
\newcommand{\diag}{\operatorname{diag}}
\newcommand{\tr}{\operatorname{tr}}
\newcommand{\Pa}{\mathcal{P}}
\newcommand{\Prod}{\mathscr{P}}
\newcommand{\Eq}{\mathord{=_2}}
\newcommand{\SEq}{\mathord{=^-_2}}
\newcommand{\NEq}{\mathord{\neq_2}}
\newcommand{\SNEq}{\mathord{\neq^-_2}}
\newcommand{\EQ}{\mathcal{E}Q}
\newcommand{\supp}{\operatorname{supp}}

\newcommand{\Holant}{\mathrm{Holant}}
\newcommand{\CSP}{\mathrm{\#CSP}}
\newcommand{\pij}[2]{\partial_{#1}^{#2}}

\title{From a finite subgroup of $SO(3)$: $C_2$ leads to a Holant dichotomy}
\author{Yuxuan Zhou\thanks{Key Laboratory of System Software (Chinese Academy of Sciences), Beijing, China}\textsuperscript{\phantom{*},}\thanks{State Key Laboratory of Computer Science, Institute of Software, Chinese Academy of Sciences, Beijing, China}\textsuperscript{\phantom{*},}\thanks{University of Chinese Academy of Sciences, Beijing 100080, China}\\
	\href{mailto:zhouyx@ios.ac.cn}{\texttt{zhouyx@ios.ac.cn}}}
\date{}
\begin{document}
\maketitle
\begin{abstract}
We prove an FP versus \#P-hardness dichotomy for Boolean Holant problems defined by finite sets of even-arity signatures with algebraic complex values, whose nonzero binary gadget signatures form two nonsingular projective classes, the cyclic group $C_2$ of order two. Transpose closure gives two orthogonally distinct forms: signed equality and skew-symmetric disequality.

For decomposable quaternary gadgets, we combine phase sharing, full-rank realnumberization and factor integration. In one isotropic branch, a tail obstruction forces common one-sided support; weight conservation then removes noncentral layers and the surviving tensors become real. An eight-port Reed--Muller resource handles the other residual branch. Unavailable tensor factors can enlarge the binary group; we classify these extensions directly. In the nonstandard four-class extension, exact six- and eight-port base cases and a general factor-propagation argument force every signature to be a binary tensor product, giving polynomial-time evaluation.

Indecomposable quaternary gadgets lead, through interpolation and balanced equalities, to \#CSP dichotomies applied to the entire transformed signature set. In the balanced isotropic case, non-affine support produces an ordinary \#CSP obstruction or an encoded hard six-vertex model. Affine balanced supports admit fixed opposite pairings; the resulting \#CSP classification gives the common affine or product FP cases and hardness otherwise. The group hypothesis is imposed only on the original gadget closure, not on signature sets obtained merely by Turing equivalence.
\end{abstract}
\newpage
\section{Introduction}
A Boolean Holant problem assigns a fixed local function, called a signature, to each vertex of an input graph. Each edge receives a bit. For one assignment of bits to edges, multiply the values of the vertex signatures on their incident bits; then sum over all assignments. The resulting sum is the Holant value. For example, if each vertex accepts exactly the assignments with one incident bit equal to one, the sum counts perfect matchings. Complex-valued signatures allow weights and cancellations. Given a fixed finite signature set $\F$, the classification problem asks whether these sums can be evaluated efficiently or whether $\Holant(\F)$ is \#P-hard.

Several major cases are already classified. The real Boolean Holant dichotomy of Shao and Cai~\cite{shao2020dichotomy} states that every fixed finite real-valued signature set gives an FP or \#P-hard problem; we state the result formally in Theorem~\ref{thm:real-holant}. Classifications also hold for symmetric signatures and for several frameworks with auxiliary unary signatures~\cite{cai2013complete,cai2011dichotomy,backens2018complete}. For complex weights, the odd-arity dichotomy of Meng, Wang, Xia and Zheng~\cite{meng2025odd} handles every signature set containing a nonzero odd-arity signature. Its upper bound is $\mathrm{FP}^{\mathrm{NP}}$, inherited from weighted Eulerian orientations~\cite{meng2025eo}. Here FP denotes polynomial-time exact computation, and $\mathrm{FP}^{\mathrm{NP}}$ allows a polynomial-time algorithm an NP oracle. These results leave even-arity complex-valued signature sets as the remaining setting considered in this paper.

Binary signatures provide a useful way to organize that setting~\cite{xia2026framework}. A binary signature is a $2\times2$ matrix. Connecting binaries in a path multiplies their matrices; reversing the two ports transposes the matrix. A gadget is a finite network with some ports left open, so its boundary also defines a signature. We identify two nonzero binary gadget signatures when their matrices differ by a nonzero scalar. If there are finitely many such classes and all are nonsingular, composition makes them a finite group.

We study $C_2$, the cyclic group of order two: $C_2=\{1,s\}$ with $s^2=1$. Thus the hypothesis is that the original gadget closure has exactly two nonzero binary classes, both nonsingular. Our main result is the following.

\noindent\textbf{Main result (Theorem~\ref{thm:main}).} Let $\F$ be a fixed finite set of Boolean signatures of even arity with algebraic complex values. If its nonzero binary gadget signatures form exactly two nonsingular projective classes, then $\Holant(\F)$ is in FP or is \#P-hard. The group condition is imposed on actual gadgets over $\F$.

The proof has two independent case distinctions. Transpose closure first gives the normal classes $[\Eq],[\SEq]$, where $\SEq=\diag(1,-1)$, or the isotropic classes $[\Eq],[\SNEq]$, where $\SNEq=\left(\begin{smallmatrix}0&1\\-1&0\end{smallmatrix}\right)$. The normal form is obtained by an orthogonal change of basis, which preserves equality edges. Within each case, either every nonzero quaternary gadget is tensor-decomposable, or an indecomposable one exists. Decomposability has its ordinary tensor meaning; the factors need not be available gadgets. Table~\ref{tab:proof-overview} records the resulting four branches.

\begin{table}[H]
\centering
\small
\begin{tabular}{@{}>{\raggedright\arraybackslash}p{0.13\textwidth}
>{\raggedright\arraybackslash}p{0.40\textwidth}
>{\raggedright\arraybackslash}p{0.40\textwidth}@{}}
\toprule
Binary case & Every quaternary gadget decomposes & An indecomposable quaternary exists \\
\midrule
Normal & Phase sharing or a binary-product extension
(Corollary~\ref{thm:normal-decomp-all}) &
Interpolation to even-occurrence \#CSP
(Theorem~\ref{thm:tnm}) \\
\addlinespace
Isotropic & Factor integration, one-sided realnumberization, and a
Reed--Muller resource (Theorem~\ref{thm:iso-complete}) &
Balanced equality, support structure, and \#CSP reductions
(Theorem~\ref{thm:iso-trueq}) \\
\bottomrule
\end{tabular}
\caption{Proof organization under the original gadget-$C_2$ hypothesis.
Each branch classifies the corresponding Holant problem in FP or as \#P-hard.}
\label{tab:proof-overview}
\end{table}

The main tool in the decomposable branch is realnumberization: in a common basis, remove one nonzero scalar from each signature so that every remaining entry is real. The real Holant dichotomy then applies. To prove this, pair all ports of a signature and contract them with available binaries. Changing two pairs at a time leaves a quaternary residual tensor. Its controlled binary factors allow the pairing and contraction labels to change while preserving a nonzero value up to a real ratio. This is the sharing lemma. When these real linear tests have full rank, their common phase forces the signature itself to be real up to a scalar.

Two obstacles require further arguments. First, an unavailable tensor factor cannot be used as though it were an original gadget. In the normal case, an antidiagonal factor supplies an actual tensor square. Pairing its occurrences proves the structure of a four-class extension independently of the reduction that adjoins the factor. The real Pauli extension permits realnumberization; the nonstandard extension contains only binary tensor products and is evaluated by matrix products around cycles. Second, in isotropic coordinates the original contraction tests see only the half-weight layer. A minimum-arity argument integrates its binary factors. A remaining single extreme tail gives a polynomial family of actual gadgets. Any input with support in the opposite weight direction would make one of these gadgets realize a forbidden binary. The resulting common support direction makes every noncentral contribution vanish in a closed instance. The middle restrictions have a shared phase and become real by full rank on their conjugation-invariant subspace. The real Holant dichotomy then gives an FP versus \#P-hardness classification. At arity eight a Reed--Muller exception supplies additional tests, with a common phase established by padding with the same real resource. The six- and eight-port classifications follow from the finite contraction identities below; the factor-integration argument covers every larger arity.

In the indecomposable branch, binary contractions restrict the complementary entries of a quaternary gadget. Two-copy gadgets and a common projector interpolation then produce equalities, except for two residual block forms. For the projector used in the isotropic case, polynomial identities prove that its extension still has the two binary classes. A balanced equality makes every balanced support closed under complementation. Non-affine support produces either a non-affine ordinary \#CSP restriction or a hard six-vertex model encoded by parallel bundles. Affine balanced supports have fixed opposite pairings, so the whole Holant problem is equivalent to ordinary \#CSP on the compressed signatures. Its FP alternatives require all compressed signatures to belong to one common affine or product class. Noncentral inputs lead to weighted-equality \#CSP reductions or to the preceding one-sided restriction. The residual block forms give the same one-sided reduction or an equality after a common holographic transformation. These reductions retain every input signature and give FP versus \#P-hardness.

Our proof builds on factor integration and opposite-pair encoding from the Eulerian-orientation classification of Cai, Fu and Shao~\cite{cai2020eoars}, and on bounded-occurrence replacement from the real Holant proof of Shao and Cai~\cite{shao2020dichotomy}. Here the original $C_2$ condition and the quaternary structure supply the phase and support constraints needed for complex-valued signatures. We prove the shared phase before applying the real dichotomy. We also prove that the isotropic projector interpolation preserves the two binary classes, which permits the subsequent support analysis and whole-set \#CSP classification.

The finite-group viewpoint organizes the possibilities; the present result addresses only $C_2$. It does not classify the trivial group, larger cyclic groups, dihedral or polyhedral groups, or infinite binary closures. Fewer binaries can leave too few contraction tests, while more binaries require new control of the residual factors and their phases. Moreover, identifying a finite projective matrix group with a subgroup of $\mathrm{SO}(3,\mathbb R)$ does not in general supply an orthogonal holographic reduction. Section~\ref{sec:3} distinguishes these operations and proves the required $C_2$ normal forms directly.

Section~\ref{sec:prel} fixes the notation and states the imported classification results. Section~\ref{sec:3} states the main theorem. Sections~\ref{sec:4} and~\ref{sec:5} prove the decomposable and indecomposable branches, respectively.

\section{Preliminaries}
\label{sec:prel}
\subsection{Definitions and notations}
A Boolean signature of arity \(r\) is a function \(f:\{0,1\}^r\to\mathbb{C}\). We call it real-valued if all its entries lie in \(\mathbb{R}\). Throughout this paper, all input signatures take algebraic values, and every input signature set is fixed and finite. All exact arithmetic is carried out in a fixed algebraic number field, which is enlarged once to contain the constants arising from any fixed change of basis or reduction. Graphs may have parallel edges and self-loops, and the incident ports at each vertex are ordered. Identically zero signatures and scalar components can be treated separately.

The support of a signature $f$ of arity $r$ is
\[
\supp(f)=\{x\in\{0,1\}^r:f(x)\ne0\}.
\]

For $s\in\{0,1\}^r$, write $s_i$ for its $i$th bit, $|s|$ or $\HW(s)$ for its Hamming weight, and $\bar s$ for its bit-wise complement. Thus $|s|+|\bar s|=r$. We also write $f_s=f(s)$. A signature has even parity if it vanishes on all odd-weight inputs, and odd parity if it vanishes on all even-weight inputs. It has definite parity if one of these conditions holds. Arity counts the number of variables; parity restricts the Hamming weights in the support. For strings of length $r$, define
\[
\mathrm{HW}^{\ge}_r=\{s:|s|\ge r/2\},\qquad
\mathrm{HW}^{\le}_r=\{s:|s|\le r/2\},\qquad
\mathrm{HW}^{=}_r=\{s:|s|=r/2\}.
\]
We omit the subscript when the arity is clear. A signature is $\mathrm{HW}^{\ge}$, $\mathrm{HW}^{\le}$, or $\mathrm{HW}^{=}$ when its support lies in the indicated set. The last condition is also called balanced or Eulerian-orientation (EO) support. A common one-sided condition on a signature set uses the same inequality for every signature. A signature is symmetric if its value depends only on Hamming weight; in that case $[f_0,\ldots,f_r]$ denotes its values on the $r+1$ weight layers. A nonzero signature is tensor-decomposable if, after a permutation of its variables, it is a tensor product of two positive-arity signatures. Otherwise it is tensor-indecomposable. Assertions that a family of gadgets is decomposable concern its nonzero members; zero gadgets are handled separately.

\begin{definition}[Holant problems]
\label{def:hol}
A Holant problem $\Holant(\F)$ is parameterized by a set $\mathcal{F}$ of signatures $f:\{0,1\}^r\to\mathbb{C}$. An instance $I$ is a signature grid $\Omega=(G,\pi)$, where $G=(V,E)$ is a graph and $\pi$ assigns to each vertex $v\in V$ a signature $f_v\in\mathcal{F}$ of arity $\deg(v)$. The output is the partition function of $\Omega$ (also called Holant value, denoted by $Z(I)$)
\[
\mathrm{Holant}_{\Omega}= \sum_{\sigma:E\to\{0,1\}}\prod_{v\in V} f_v\bigl(\sigma|_{E(v)}\bigr),
\]
where $E(v)$ is the ordered list of incidences at $v$, so a self-loop appears twice, and $\sigma|_{E(v)}$ gives one bit at each port in that order. We also write $\mathrm{Holant}(\Omega)$ for $\mathrm{Holant}_{\Omega}$. The bipartite Holant problem $\mathrm{Holant}(\mathcal{F}\mid\mathcal{G})$ is defined on bipartite graphs $H=(U,V,E)$, with $U$ assigned by $\mathcal{F}$ and $V$ by $\mathcal{G}$. A bipartition with these prescribed label sets is part of that problem's definition.
\end{definition}

Multiplying a signature by a fixed nonzero scalar preserves complexity. If $f$ occurs $m$ times, replacing it by $\lambda f$ multiplies the instance value by the known factor $\lambda^m$.

We will use $\leq_T$ and $\equiv_T$ to represent the polynomial-time Turing reductions and equivalences. Write $\langle\F\rangle$ for the gadget closure of $\F$, including permutations and disconnected tensor products. We use $\B(\F)$ for its binary signatures, up to nonzero scalar. The main theorem's group condition concerns this gadget closure. The problem $\Holant(\F)$ permits all graphs. Its bipartite presentations below are obtained by explicit equivalences, starting with edge subdivision; they impose no additional restriction on the original input graphs. A Turing reduction adjoining $g$ does not establish $g\in\langle\F\rangle$. Whenever a reduction enlarges the input signature set, any structural claim about its new gadget closure is proved separately. Computational problems always have finite input signature sets. Only finitely many fixed gadgets are adjoined in any reduction, unless a polynomial-size interpolation is explicitly described. When listing $\B(\F)$, we choose one representative of each nonzero scalar class. Since $\Holant(\F) \equiv_T \Holant(\F\cup \{=_2\})$, we can always assume that $=_2\in \F$. Section~\ref{sec:3} analyzes the projective matrix structure of $\B(\F)$.
\subsection{Fundamental methods}
\subsubsection{Signature matrix and gadget construction}
Let \(f\) be a signature of arity \(n\) and fix an ordering of its variables. For a chosen subset \(S_k\) of \(k\) variables, we represent \(f\) as a \(2^k \times 2^{n-k}\) matrix, denoted \(M_{S_k}(f)\), whose rows are indexed by the lexicographically ordered assignments to the variables in \(S_k\) (from \(\mathbf{0}_k\) to \(\mathbf{1}_k\)), and whose columns are indexed by the lexicographically ordered assignments to the remaining \(n-k\) variables. The entry at a given row and column is the value of \(f\) on the input string obtained by concatenating the row assignment and the column assignment according to the fixed variable ordering.

When \(S_k\) consists precisely of the first \(k\) variables \(x_1,\dots,x_k\) in the ordering, we write the matrix simply as \(M(f)_{x_1\cdots x_k,\;x_{k+1}\cdots x_n}\), or simply \(M_f\) when no confusion arises. For signatures of even arity $n=2d$, one often takes \(k = n/2 = d\).

The Pauli matrices provide useful examples of binary signatures:
$$
\sigma_x = \begin{pmatrix} 0 & 1 \\ 1 & 0 \end{pmatrix},\quad 
\sigma_y = \begin{pmatrix} 0 & -i \\ i & 0 \end{pmatrix},\quad 
\sigma_z = \begin{pmatrix} 1 & 0 \\ 0 & -1 \end{pmatrix}.
$$
In the study of isotropic binary signatures and Bell matrices, the Pauli matrices together with the identity play a fundamental role. We use the following representatives, obtained by multiplying the three Pauli matrices by $i$:
$$
 X = i\sigma_x =\begin{pmatrix}0&i\\i&0\end{pmatrix} ,\quad
 Y = i\sigma_y = \begin{pmatrix}0&1\\-1&0\end{pmatrix},\quad
 Z = i\sigma_z = \begin{pmatrix}i&0\\0&-i\end{pmatrix}.
$$
Together with the identity matrix $I$, these form $\Pa=\{I,X,Y,Z\}$. The corresponding binary signatures are
\[
\Eq=(1,0,0,1),\quad \SEq=(1,0,0,-1),\quad
\NEq=(0,1,1,0),\quad \SNEq=(0,1,-1,0).
\]
Thus their matrices satisfy
\[
M(\Eq)=I,\quad M(\SEq)=-iZ,\quad
M(\NEq)=-iX,\quad M(\SNEq)=Y.
\]
We use $\NEq$ also for unsigned disequality $\neq^+_2$. In matrix products and local transformations, a binary signature denotes its matrix; for example, $(\SEq)^2=I$. The notation $(\NEq)_{uv}$ means $\NEq(x_u,x_v)$, while $(\SEq)_u$ denotes the matrix acting on port $u$ alone. These four real signatures form a basis of binary tensors. A basis element is an available gadget only when its realizability has been established.

These matrices will be used throughout the rest of the paper. When there is no ambiguity, we simply write \(I, X, Y, Z\) for the corresponding matrices or signatures; in particular, \(I\) denotes the identity matrix (the \(2\times 2\) identity unless stated otherwise). The matrices above fix the Pauli notation used in the following sections.

Gadget construction is the fundamental reduction technique in the study of Holant and related counting problems. It allows us to build new signatures from a given set \(\mathcal{F}\) by connecting existing signatures via shared variables (edges). The resulting signatures are called \emph{realizable}, and adding them to the signature set does not change the computational complexity of the Holant problem. We formalize this construction as an $\F$-gate.

\begin{definition}
Let \(\mathcal{F}\) be a set of signatures. An \(\mathcal{F}\)-gate is a pair \((G,\pi)\) where \(G = (V,E,D)\) is a graph whose edges are partitioned into two types:
\begin{itemize}
\item Internal edges \(E\): both endpoints lie in \(V\);
\item Dangling edges \(D\): exactly one endpoint lies in \(V\), the other end being free.
\end{itemize}
    
Vertices in \(V\) are assigned signatures from \(\mathcal{F}\) as in a signature grid. Let \(|E| = n\) and \(|D| = m\). Identify the internal edges with variables \(\{x_1,\dots,x_n\}\) and the dangling edges with variables \(\{y_1,\dots,y_m\}\). For a fixed assignment \((y_1,\dots,y_m)\in\{0,1\}^m\) to the dangling edges, the gate evaluates to
\[
f(y_1,\dots,y_m) \;=\; \sum_{\sigma: E\to\{0,1\}} \;\prod_{v\in V} f_v\bigl(\hat\sigma|_{(E\cup D)(v)}\bigr),
\]
where $\hat\sigma:E\cup D\to\{0,1\}$ extends $\sigma$ by the given dangling assignment, and $(E\cup D)(v)$ is the ordered list of all internal and dangling incidences at $v$, again counting both ports of a self-loop. The function $f$ is the {signature realized by the gate}. If such an $\F$-gate exists, $f$ is {realizable} from $\F$ via gadget construction.

Moreover, when there are no internal edges (\(E = \emptyset\)), i.e., the realized signature is simply the tensor product of the signatures on all vertices, the signature can be denoted as \(\bigotimes_{v\in V} f_v\).
\end{definition}
A fundamental property \cite{cai2017complexity} states that if \(f\) is realizable from \(\mathcal{F}\), then
\[
\mathrm{Holant}(\mathcal{F}) \equiv_T \mathrm{Holant}(\mathcal{F}\cup\{f\}).
\]
This equivalence justifies adjoining a fixed gadget. Whenever a proof below refers to a realizable signature as available, it uses this equivalence; the underlying input signature set remains fixed and finite. Structural hypotheses and arguments apply to the entire gadget closure $\langle\F\rangle$. Thus we work with a gadget-closed collection for structural purposes, while $\Holant(\F)$ remains the problem defined by its finite set of input signatures.

We use two basic gadget operations. The first closes two ports through a binary signature and is called contraction. Let \(f\) be a signature of even arity \(2d\) (the construction works for any arity, but the typical case is even). Choose any two distinct variables of \(f\), say \(x_i\) and \(x_j\). Introduce a new vertex assigned a binary signature \(b\) (a \(2\times 2\) matrix), and connect \(x_i\) and \(x_j\) to the two dangling edges of \(b\). The resulting gadget has arity \(2d-2\) and yields a new signature. Following previous works~\cite{shao2020dichotomy}, we denote the contraction operation on variables $x_i$ and $x_j$ using the binary signature $b$ as $\partial_{ij}^{b}f$. In matrix form, if we order the variables so that \(x_i\) and \(x_j\) are the first two, the operation corresponds to
\begin{equation}\label{eq:contraction}
\partial_{ij}^{b}f(\mathbf{x}) = b_{00}f(0,0,\mathbf{x}) + b_{01}f(0,1,\mathbf{x}) + b_{10}f(1,0,\mathbf{x}) + b_{11}f(1,1,\mathbf{x}),
\end{equation}
where \(\mathbf{x}\) denotes the remaining variables. We use this construction with signed or unsigned equality and disequality signatures. Up to their specified nonzero scalars, these are the contractions $\partial^P_{ij}f$ for $P\in\Pa$.

A more powerful construction, sometimes called the \emph{mating gadget}, combines two signatures by identifying a matching of their variables. Let \(f\) and \(g\) be signatures of arities \(2d_1\) and \(2d_2\), respectively (the evenness is assumed for convenience). Choose an integer \(m\) with \(0 \le m \le \min(2d_1, 2d_2)\) and select \(m\) distinct variables from \(f\) and \(m\) distinct variables from \(g\). Identify these selected variables pairwise (i.e., connect the \(k\)-th selected variable of \(f\) with the \(k\)-th selected variable of \(g\) via an internal edge). The remaining \(2d_1 - m\) variables of \(f\) and \(2d_2 - m\) variables of \(g\) become dangling edges of the combined gadget, so the resulting signature \(h\) has arity \(2(d_1 + d_2 - m)\).

In the signature matrix representation (with variables ordered appropriately), the mating gadget corresponds precisely to matrix multiplication. Specifically, arrange the variables so that the \(m\) shared (matched) variables appear as the last \(m\) variables of \(f\) and the first \(m\) variables of \(g\). Then the matrix of \(f\) can be viewed as a \(2^{2d_1 - m} \times 2^{m}\) matrix, and the matrix of \(g\) as a \(2^{m} \times 2^{2d_2 - m}\) matrix. The resulting signature \(h\) has matrix
\[
M_h = M_f \cdot M_g,
\]
which is a \(2^{2d_1 - m} \times 2^{2d_2 - m}\) matrix. This operation generalizes the familiar composition of linear maps and is the algebraic backbone of many reductions in the Holant framework.

\subsubsection{Holographic transformations}
\label{sssec:holographic-transformations}
Holographic transformations are naturally formulated on bipartite graphs. Subdividing each edge once and assigning equality \(=_2\) to the new vertex makes the graph bipartite without changing its Holant value. Thus \(\operatorname{Holant}(=_2\mid\mathcal F) \equiv_T\operatorname{Holant}(\mathcal F)\).

For an invertible \(2\times 2\) matrix \(T\in\mathbf{GL}_2(\mathbb{C})\) and a signature \(f\) of arity \(n\) (written as a column vector), the transformed signature is \(T^{\otimes n}f\). For a set \(\mathcal{F}\), write \(T\mathcal{F} = \{T^{\otimes n}f \mid f\in\mathcal{F}\}\). On the left, signatures are regarded as row tensors and multiplied by $(T^{-1})^{\otimes n}$; equivalently, left column tensors are multiplied by $(T^{-T})^{\otimes n}$. In particular, a left binary kernel $B$ becomes $T^{-T}BT^{-1}$. For a right-hand binary $f$, the corresponding formula is $M(Tf)=T M(f)T^T$. These are ordinary transposes: tensor contraction is bilinear and does not conjugate entries.

\begin{theorem}[\cite{valiant2008Holographic}]
\label{t:holo}
For any $T \in \mathbf{GL}_2(\mathbb{C})$, $\Holant(\F \mid \G) \equiv_T \Holant(\F T^{-1}\mid T\G)$.
\end{theorem}

As an important special case, consider the matrix
\[
 K = \frac{1}{\sqrt{2}}\begin{pmatrix} 1 & 1 \\ i & -i \end{pmatrix},\qquad
 K^{-1} = \frac{1}{\sqrt{2}}\begin{pmatrix} 1 & -i \\ 1 & i \end{pmatrix}.
\]
The identity $K^TIK=\NEq$ transforms the left equality kernel into disequality when $T=K^{-1}$. Thus
\[
\operatorname{Holant}(=_2 \mid \mathcal{F}) \equiv_T \operatorname{Holant}(\neq_2 \mid K^{-1}\mathcal{F}).
\]
For an arity-$r$ right-hand signature, write $\hat f=(K^{-1})^{\otimes r}f$ and $\hat\F=K^{-1}\F$. For a binary signature this means $M(\hat f)=K^{-1}M(f)(K^{-1})^T$, giving the explicit correspondences
\[
\Eq\longmapsto \NEq,\qquad \SEq\longmapsto \Eq,\qquad
\NEq\longmapsto -i\SEq,\qquad \SNEq\longmapsto i\SNEq.
\]
Thus the four projective classes are permuted. Their scalar factors are known and are retained whenever the signatures are used in a reduction.

Moreover, orthogonal matrices interact well with this type of transformation. Let \(\Oth\) denote the set of orthogonal matrices of order $2$, i.e.
\[
\Oth = \{\, M \in \mathbf{GL}_2(\mathbb{C}) \mid M^{T} M = I \,\},
\]  
where $I$ is the $2\times2$ identity. For $Q\in\Oth$, $Q^{-T}IQ^{-1}=I$, so the left equality kernel is preserved and $\Holant(\F)\equiv_T\Holant(Q\F)$. The transformed right-hand binary matrices are $Q M(f)Q^T$.

\subsubsection{Polynomial interpolation}

Polynomial interpolation is a powerful tool for constructing reductions. A detailed treatment can be found in~\cite[Section~1.3.3]{cai2017complexity}.

The following standard lemma illustrates binary interpolation. We then give the quaternary projector interpolation used in Section~\ref{sec:5}.

\begin{lemma}[{\cite[Lemma~2.26 of the full version]{shao2020dichotomy}}]
Let \(f\) and \(g\) be nonzero binary signatures with
\[
M(f)=P^{-1}\begin{pmatrix}\lambda_1&0\\0&\lambda_2\end{pmatrix}P,\qquad 
M(g)=P^{-1}\begin{pmatrix}1&0\\0&0\end{pmatrix}P
\]  
for some invertible matrix \(P\). If \(\lambda_1\neq0\) and \(\left|\frac{\lambda_2}{\lambda_1}\right|\neq1\), then
\[
\mathrm{Holant}(g,\mathcal{F})\le_T\mathrm{Holant}(f,\mathcal{F})
\]  
for any signature set \(\mathcal{F}\).
\end{lemma}
Related interpolation constructions also apply in planar settings~\cite{cai2017holographic}. For the unrestricted graphs considered here, the following direct argument suffices.
\begin{lemma}
\label{lem:intp}
Suppose $\F$ contains a quaternary signature $f$ with
\[
M(f)=\diag(a,b,c,d),\qquad ad-bc\ne0.
\]
Then $\Holant(\F,\{=_4\})\leq_T\Holant(\F)$.
\end{lemma}
\begin{proof}
Swap the second and third ports of $f$. The resulting signature $g$ has the block $B=\left(\begin{smallmatrix}a&b\\c&d\end{smallmatrix}\right)$ on $W=\operatorname{span}\{00,11\}$ and is zero on the other two coordinates. By hypothesis $B$ is invertible. A chain of $k\ge1$ copies of $g$ has block $B^k$ on $W$ and zero elsewhere.

Consider an instance with $m\ge1$ desired occurrences of $=_4$. Their identity blocks on $W$ give the target value $Z_0$; replacing each by a length-$k$ chain gives a value $Z_k$ that is the same linear functional applied to $(B^{\otimes m})^k$. Let $\lambda,\mu\ne0$ be the eigenvalues of $B$, counted with multiplicity. The polynomial
\[
p(z)=\prod_{j=0}^{m}(z-\lambda^j\mu^{m-j})^{m+1}
     =\sum_{k=0}^{D}p_kz^k
\]
annihilates $B^{\otimes m}$, and $p_0\ne0$. Indeed, a tensor product of $m$ Jordan blocks of size at most two has Jordan blocks of size at most $m+1$: after removing its eigenvalue, each multiplication raises total degree in commuting nilpotent variables whose squares are zero. Applying the linear functional to this identity gives
\[
Z_0=-p_0^{-1}\sum_{k=1}^{D}p_kZ_k.
\]
Here $D\le(m+1)^2$, so all queries and chains have polynomial size. The case $m=0$ needs no replacement. Construct $p$ from the two fixed eigenvalues; the exponentially large matrix $B^{\otimes m}$ is used only in the proof, not formed by the algorithm. The coefficients of $p$ have polynomial bit length in a fixed algebraic number field. This proof covers zero entries of $B$, repeated eigenvalues, and roots of unity.
\end{proof}

An analogous projector interpolation holds for an invertible middle block: assume $ad-bc\ne0$. For a signature $f\in\F$ with
$M(f) = \begin{pmatrix}
     0&0&0&0\\
     0&a&b&0\\
     0&c&d&0\\
     0&0&0&0
\end{pmatrix}$,
we can obtain $M(g) = \begin{pmatrix}
     0&0&0&0\\
     0&1&0&0\\
     0&0&1&0\\
     0&0&0&0
\end{pmatrix}$ via the same technique, i.e. $\Holant(\F,g,f)\leq_T\Holant(\F,f)$.

For another interpolation example, write $\neq_2^{\,a,b}$ for the binary signature with matrix
$\begin{pmatrix}0&a\\b&0\end{pmatrix}$.  In the bipartite setting
$\operatorname{Holant}(\neq_2\mid\cdot)$, connecting $k$ copies of $\neq_2^{\,1,x}$ in a path realizes $\neq_2^{\,1,x^{k}}$: if $B(x)=\left(\begin{smallmatrix}0&1\\x&0\end{smallmatrix}\right)$ and $\NEq=\left(\begin{smallmatrix}0&1\\1&0\end{smallmatrix}\right)$, then $B(x)\NEq=\diag(1,x)$ and the path matrix is $(B(x)\NEq)^k\NEq$.

\begin{lemma}\label{lem:interp}
Let $\F$ be a signature set and let $x\neq0$ be a complex number that is not a root of unity. If $\neq_2^{\,1,x}$ is realizable over $\F$ with native $\NEq$ edges, the rank-one binary $\Delta=\neq_2^{\,1,0}$ can be adjoined:
\[
\Holant(\NEq\mid\F,\Delta)\leq_T\Holant(\NEq\mid\F).
\]
\end{lemma}
\begin{proof}
Let $I$ be an instance of $\operatorname{Holant}(\neq_2\mid\F\cup\{\Delta\})$ in which $\Delta$ occurs $m$ times.  Replace each $\Delta$ by the variable signature $\neq_2^{\,1,a}$.  The resulting partition function is a polynomial
\[
Z(I,a)=\sum_{j=0}^{m}c_j a^{j},
\]
because the $m$ replaced signatures contribute exactly $j$ factors of $a$ to a term in which $j$ of their inputs equal $10$.  In particular, $Z(I,0)$ is the original partition function.

For $j=1,\dots,m+1$, connect $j$ copies of $\neq_2^{\,1,x}$ in a path to realize $\neq_2^{\,1,x^{j}}$, replace each $\Delta$ by this signature, and compute $Z(I,x^{j})$.  This gives the Vandermonde system
\[
\sum_{\ell=0}^{m}c_\ell\,(x^{j})^{\ell}=Z(I,x^{j}),
\qquad j=1,\dots,m+1.
\]
Since $x$ is not a root of unity, the values $x,x^2,\dots,x^{m+1}$ are distinct, so the system is invertible.  Solving it in polynomial time recovers $c_0$, hence $Z(I,0)$.
\end{proof}

\subsection{Previous works}

We collect the complexity classifications and reduction lemmas used in the subsequent proofs.

\subsubsection{Dichotomy for Holant problems with real values, and realnumberization}
In a fixed basis, realnumberization means choosing a nonzero scalar $\lambda_f$ for each nonzero input $f$ so that $\lambda_f f$ is real-valued. The instance's product of scalar factors is known, so this operation preserves complexity. Lemma~\ref{lem:real} formalizes this reduction using the following theorem.
\begin{theorem}[Dichotomy for Holant problems with real values~\cite{shao2020dichotomy}]
\label{thm:real-holant}
Let $\F$ be a fixed finite set of real-valued Boolean signatures. Then $\Holant(\F)$ is in FP or is \#P-hard.
\end{theorem}
\subsubsection{\#CSP problems}
We give the definition of the framework of \#CSP, which is closely related to Holant \cite{cai2017complexity}. A variant of \#CSP, known as \#CSP$_d$ \cite{cai2017complexity,meng2025odd}, also plays a role in our proof.

\begin{definition}[\#CSP, \cite{cai2017complexity}]
Let $\mathcal{F}$ be a fixed finite set of complex-valued signatures over the Boolean domain. An instance of $\CSP(\mathcal{F})$ consists of a finite variable set $X = \{x_1, x_2, \ldots, x_n\}$ and a finite collection $C$ of constraint occurrences. Each constraint specifies a signature $f\in\mathcal F$ of arity $k$ and an ordered tuple $(x_{i_1},x_{i_2},\ldots,x_{i_k})$ of variables from $X$, allowing repeated variables and repeated constraints. Its value is
\[
Z(I) = \sum_{x_1, \ldots, x_n \in \{0,1\}} \prod_{(f, x_{i_1}, \ldots, x_{i_k}) \in C} f(x_{i_1}, \ldots, x_{i_k}).
\]
\end{definition}

\begin{definition}
Let $d \ge 1$ be an integer and let $\mathcal{F}$ be a set of complex-valued signatures. The problem $\CSP_d(\mathcal{F})$ is the restriction of $\CSP(\mathcal{F})$ to the instances where every variable occurs a multiple of $d$ times.
\end{definition}

We reserve $\Pa$ for the Pauli set. For the complexity classifications below, $\Prod$ is the class of products of unary signatures and binary equality or disequality constraints, on possibly overlapping sets of variables. The class $\mathcal A$ consists of signatures
\[
f(x)=c\,\mathbf1_{Lx=b}\,i^{Q(x)},
\]
where $Lx=b$ is an affine system over $\mathbb F_2$, and $Q$ is a polynomial modulo four of degree at most two with even coefficients on its mixed quadratic terms. Both classes include zero. In particular, every nonzero member of either class has affine support.
\begin{theorem}[Boolean \#CSP dichotomy]\label{thm:csp-standard}
For a finite Boolean signature set $\G$, $\CSP(\G)$ is in FP if $\G\subseteq\mathcal A$ or $\G\subseteq\Prod$, and is \#P-hard otherwise.
\end{theorem}
We use the formulation in Theorem 18 of~\cite{meng2025odd}; the same theorem is stated as Theorem 2.3 of~\cite{cai2018complexity6}.

For an integer $d \ge 1$, denote by $\EQ_d$ the class of equality signatures whose arities are divisible by $d$:
\[
\EQ_d = \{=_d, =_{2d}, \ldots, =_{rd}, \ldots\},
\]
where $=_d$ is a symmetric signature: $(=_d) = [1,0,\cdots,1]$. Sometimes $\EQ_1$ will be denoted as $\EQ$, which contains all equality signatures.
\begin{lemma}[{\cite[Lemma~1.2 and Eq.~(1.6)]{cai2017complexity}}]
\[
\CSP(\mathcal{F}) \equiv_T \Holant(\EQ \cup \mathcal{F}),
\]
Moreover, for an integer $d \ge 1$,
\[
\CSP_d(\mathcal{F}) \equiv_T \Holant(\EQ_d \mid \mathcal{F}),
\]
In the bipartite representation, each variable with positive occurrence count is an equality vertex; each isolated variable contributes a separately computed factor of two.
\end{lemma}
For $\CSP_2$, quaternary equality already suffices.
\begin{lemma}
\[
\CSP_2(\mathcal{F}) \leq_T \Holant(\mathcal{F}\cup\{=_4\}).
\]
\end{lemma}

Every even equality of arity at least four is realized by a tree of $=_4$ vertices, and $=_2$ is an equality wire. Conversely, in an instance over $\F\cup\{=_4\}$, each connected component of equality vertices identifies an even number of remaining ports: if it has $v$ vertices and $e$ internal edges, the number is $4v-2e$. Each edge directly joining two $\F$ ports becomes a variable with two occurrences. Collapsing the equality components therefore gives a $\CSP_2(\F)$ instance; a component with no remaining ports contributes the scalar two. Thus $\Holant(\F\cup\{=_4\})\equiv_T\CSP_2(\F)$. We use the even-occurrence dichotomy in this equivalent form. Its statement appears as Theorem 19 of~\cite{meng2025odd}, citing \cite{cai2018dichotomy}.
\begin{theorem}\label{thm:csp-even}
For every fixed finite Boolean signature set $\F$, $\CSP_2(\F)$ is in FP or is \#P-hard. More precisely, it is in FP when $\F$ is contained in one of $\Prod,\mathcal A,\mathcal A_2^1,\mathcal L$, and is \#P-hard otherwise.
\end{theorem}
Here $T_d=\diag(1,e^{\pi i/(2d)})$ and $\mathcal A_d^r=T_d^r\mathcal A$. A signature $f$ belongs to $\mathcal L$ if $(\bigotimes_jT_2^{x_j})f\in\mathcal A$ for every $x\in\operatorname{supp}(f)$. Thus tractability of $\CSP_2(\F)$ requires one of these common conditions on all signatures in $\F$. The theorem concerns the entire signature set, including all other signatures retained when $=_4$ is adjoined. For a fixed $d\ge3$, Theorem 20 of~\cite{meng2025odd} also gives
\[
\CSP_d(\NEq,\G)\text{ is in FP if }
\G\subseteq\Prod\text{ or }\G\subseteq\mathcal A_d^r
\text{ for some }1\le r\le d,
\]
and is \#P-hard otherwise. These are the other occurrence-restricted \#CSP dichotomies used in Section~\ref{sec:5}.

We also use the six-vertex theorem~\cite{cai2018complexity6}, Theorem 3.1: a quaternary signature supported on all six weight-two strings gives a \#P-hard problem with $\NEq$ edges. More generally, the FP cases of that theorem are $\mathcal A$, $\Prod$, and the supports with a zero in each complementary pair. The eight-vertex theorem~\cite{cai2017complexity8}, Theorem 3.1, includes even-parity quaternaries. If one extreme entry is zero, its one-signature problem equals the problem obtained by deleting both extreme entries, by weight conservation. This observation does not delete tails from other signatures in an instance with several types of vertices. Six and eight here count allowed local assignments; both models have arity four.
\subsubsection{Other conclusions}
\label{ssec:realm}

The generalized decomposition theorem~\cite{meng2025odd} either adjoins the factors of a signature by a Turing reduction or classifies the original problem through a common one-sided support condition.
\begin{lemma}[Generalized Decomposition lemma~\cite{meng2025odd}]
\label{lem:gendecomp}
Let $\mathcal F$ be a finite signature set and let $f,g$ be nonzero signatures. Then
\[
\Holant(\mathcal{F}, f, g) \equiv_T \Holant(\mathcal{F}, f \otimes g)
\]
holds unless every signature in $\hat{\mathcal{F}}'=\hat{\mathcal{F}}\cup\{\hat f\otimes\hat g\}$ is supported at or above half weight, or every signature is supported at or below half weight, with a common direction for the signature set. In that exceptional case, $\Holant(\F,f\otimes g)$ is in $\mathrm{FP}^{\mathrm{NP}}$ or is \#P-hard.
\end{lemma}

We also establish the following elementary reductions for use in our proofs. Realnumberization removes one nonzero complex scalar from each input signature so that all remaining entries are real. This preserves complexity and allows the real Holant dichotomy~\citep{shao2020dichotomy} to be applied.

\begin{lemma}
Let $\F$ be a finite signature set. If for each $f\in\F$ there is a nonzero scalar $\lambda_f$ such that $\lambda_f f$ is real-valued, then $\Holant(\F)$ is in FP or is \#P-hard.
    \label{lem:real}
\end{lemma}
\begin{proof}
Put $\F'=\{\lambda_f f:f\in\F\}$. Replacing each occurrence of $f$ multiplies an instance value by $\lambda_f$, so the total factor is known and nonzero. Since the signature set is finite, this gives $\Holant(\F)\equiv_T\Holant(\F')$. If $f\ne0$ is real up to a scalar and $f(x)\ne0$, every ratio $f(y)/f(x)$ for $y\in\supp(f)$ is real, so division by $f(x)$ suffices. Theorem~\ref{thm:real-holant} now applies to $\F'$.
\end{proof}
The next reduction contracts maximal binary paths to obtain a bipartite Holant instance.
\begin{lemma}\label{ref:bind}
Let $\F$ be a fixed finite set of even-arity signatures. Suppose its nonzero realizable binary signatures form finitely many classes up to nonzero scalar multiplication. Let $\B(\F)$ contain one representative of each such class, with $\Eq$ chosen for the equality class, and put $\F_{\ge4}=\{f\in\F:\operatorname{arity}(f)\ge4\}$. Then
\[
\Holant(\F)\equiv_T\Holant(\F_{\ge4}\mid\B(\F)).
\]
\end{lemma}
\begin{proof}
Zero signatures and nullary scalar components are evaluated separately. We may adjoin the finitely many binary representatives in $\B(\F)$ by fixed-gadget substitution and nonzero scalar normalization, retaining the scalar factors in every reduction. A bipartite instance on the right is an ordinary signature grid over the enlarged signature set on the left. Forgetting its designated bipartition and substituting the fixed gadgets for its binary labels therefore gives the right-to-left reduction.

For the reverse direction, start with any ordinary instance, whether or not its graph happens to be bipartite. A connected component consisting only of binary vertices is a cycle. Evaluate it as the trace of the ordered matrix product, with transposes recording port orientation. Every other maximal path of binary vertices lies between two ports of vertices labelled by $\F_{\ge4}$; these two ports may belong to the same vertex. Contract the path to its binary matrix, normalize to a member of $\B(\F)$, and retain the path scalar. A direct edge between two such ports is the path with matrix $I$. If a path matrix is zero, the instance value is zero.

Replace each resulting binary path by a single vertex. The new bipartition consists of the original $\F_{\ge4}$ vertices and the binary path vertices. Every edge runs between the two parts. The number of matrix multiplications and the size of the new instance are polynomial. Multiplying the separately computed scalars into its Holant value gives the original value.
\end{proof}
The construction chooses a new bipartition from the roles of the vertices. It does not require or preserve a bipartition of the original input graph.

\begin{corollary}
$\Holant(\F)\equiv_T\Holant(\F\mid\B(\F))$.
\end{corollary}
The forward reduction follows by allowing all of $\F$ on the first side of the preceding lemma. The reverse reduction again forgets the designated bipartition and substitutes the fixed binary gadgets.

\section{Classification and Main Results}
\label{sec:3}
For a nonsingular binary matrix $B$, write $[B]=\{\lambda B:\lambda\in\mathbb C\setminus\{0\}\}$. The nonzero binary gadget classes contain $[\Eq]$ and are closed under matrix multiplication and transposition. If there are finitely many classes and all are nonsingular, they form a group: repetition among the powers of any element supplies its inverse. Finiteness is needed for this argument; a general infinite matrix monoid need not contain inverses.

Following the binary-signature notation, write
\[
M(\B(\F))=\{[M(b)]: b\in\B(\F),\ b\ne0\}
\]
for this projective matrix structure. Thus $M(\B(\F))\cong C_2$ means that the original gadget closure has exactly two nonzero binary classes, both nonsingular, with the multiplication of the cyclic group of order two.

If a nonzero rank-one binary gadget occurs, its matrix is a tensor product of two nonzero unary vectors. The generalized decomposition lemma either classifies the original Holant problem in its one-sided exception, or adjoins those vectors and permits the odd-arity dichotomy. Hence the nonsingular case is the relevant remaining one for the group analysis.

Projective nonsingular $2\times2$ complex matrices form $\mathrm{PGL}_2(\mathbb C)\cong\mathrm{PSL}_2(\mathbb C)$. The determinant-one lift of a finite subgroup is finite, with kernel $\{I,-I\}$. Averaging a positive definite Hermitian form over that lift and changing to an orthonormal basis conjugates it into $\mathrm{SU}(2)$.

For clarity, the real rotation map can be specified explicitly using the Pauli matrices from Section~\ref{sec:prel}. For $v\in\mathbb R^3$, put $H(v)=-i(v_1X+v_2Y+v_3Z)$. For $U\in\mathrm{SU}(2)$ define $\rho(U)$ by
\[
U H(v)U^*=H(\rho(U)v),
\]
where $U^*$ denotes the conjugate transpose. Conjugation preserves the real space of traceless Hermitian matrices and its inner product $\tfrac12\tr(H(v)H(w))=v\cdot w$. Thus $\rho(U)$ is orthogonal; its determinant is one because $\mathrm{SU}(2)$ is connected. For a unit vector $u$, the matrix $U=\cos(\theta/2)I-i\sin(\theta/2)H(u)$ acts as rotation by $\theta$ about $u$, so $\rho$ is onto. Its kernel is $\{I,-I\}$: a matrix commuting with all three Pauli matrices is scalar, and its determinant is one. This is the double covering $\mathrm{SU}(2)\to\mathrm{SO}(3,\mathbb R)$. On complex traceless matrices, conjugation $H\mapsto SHS^{-1}$ preserves $\tfrac12\tr(H^2)$ and gives the corresponding map $\mathrm{SL}_2(\mathbb C)\to\mathrm{SO}(3,\mathbb C)$.

Consequently a finite projective binary group is abstractly isomorphic to a finite rotation subgroup. The conjugacy used to make its lift unitary need not preserve equality edges, so this observation is an organization of the possible groups, not a Holant reduction. For $C_2$, the cyclic group $\{1,s\}$ with $s^2=1$, the following direct argument supplies an orthogonal reduction. We use $B_\star$ for its nonscalar binary representative, reserving $M(f)$ for the signature matrix notation.

\begin{lemma}[The two $C_2$ normal forms]\label{lem:binary-normal-forms}
Suppose the nonzero binary gadget classes are $[\Eq],[B_\star]$ and form $C_2$. Either an orthogonal holographic transformation changes them to $[\Eq],[\SEq]$, or they are already $[\Eq],[\SNEq]$. These two cases cannot be interchanged by an orthogonal transformation.
\end{lemma}
\begin{proof}
Since $B_\star$ is nonscalar and the closure contains $B_\star^T$, we have $B_\star^T=cB_\star$ for some $c\ne0$. Transposing again gives $c^2=1$. Also $B_\star^2$ is a nonzero scalar multiple of $I$.

If $c=1$, rescale $B_\star$ so that $B_\star^2=I$. Its eigenvalues are $1,-1$: its minimal polynomial divides $(z-1)(z+1)$ and it is nonscalar. Eigenvectors for the distinct eigenvalues are orthogonal for the bilinear form $v^Tw$. Neither is isotropic, since in dimension two the orthogonal complement of an isotropic vector is its own span. Normalize the two eigenvectors to squared length one. They give a complex orthogonal change of basis taking $B_\star$ to $\SEq$ and preserving $I$.

If $c=-1$, a nonzero $2\times2$ skew-symmetric matrix is a nonzero scalar multiple of $Y=M(\SNEq)$. Finally, congruence preserves symmetry and skew-symmetry. In particular, an orthogonal change cannot turn one of these nonscalar matrices into the other.
\end{proof}

We call these the normal and isotropic cases, respectively. The terminology reflects the eigenvectors of $Y$, whose bilinear squared lengths are zero. All subsequent normal-case statements are made after the orthogonal normalization.

\begin{theorem}[The $C_2$ dichotomy]\label{thm:main}
Let $\F$ be a fixed finite Boolean signature set of even arity whose nonzero binary gadget classes are nonsingular and form $C_2$. Then $\Holant(\F)$ is in FP or is \#P-hard.
\end{theorem}
Sections~\ref{sec:4} and~\ref{sec:5} establish the four cases of the theorem. Their combination is given in Section~\ref{sec:conclusion}.

The $C_2$ hypothesis concerns the original gadget closure. A reduction that adjoins a tensor factor need not preserve that closure, and the appearance of a new binary class after a Turing equivalence is not a contradiction to this hypothesis.

Tensor decomposability has its ordinary, unrestricted meaning. The theorem places no condition on the binary classes after a Turing reduction. In particular, adjoining an unavailable tensor factor may yield a four-class signature set. Section~\ref{ssec:normal-enlargement} proves its gadget structure by explicit replacements and supplies its own classification. It does not infer a contradiction to $C_2$ from complexity equivalence alone.

\section{The decomposable quaternary signatures}
\label{sec:4}
Gadget closure is used only for structural statements; reductions adjoin fixed gadgets or explicitly polynomial-size interpolated gadgets. Here ``decomposable'' means an arbitrary nontrivial tensor decomposition.

The normal sharing proof first treats quaternary products of the available $\Eq,\SEq$ classes. Arbitrary unavailable factors are handled in Section~\ref{ssec:normal-enlargement}: the resulting four-class closure is proved by actual tensor-square replacements and then classified directly. The isotropic proof likewise checks its enlargements separately. No factor-closure assumption is added to the main theorem.

Using the realnumberization and binary-path reductions established in Section~\ref{ssec:realm}, we first prove the normal case with available binary factors by phase sharing and full-rank reconstruction. We then collect the Pauli-basis tools and complete the normal case for arbitrary factors, before treating the isotropic case.

\subsection{Normal case}
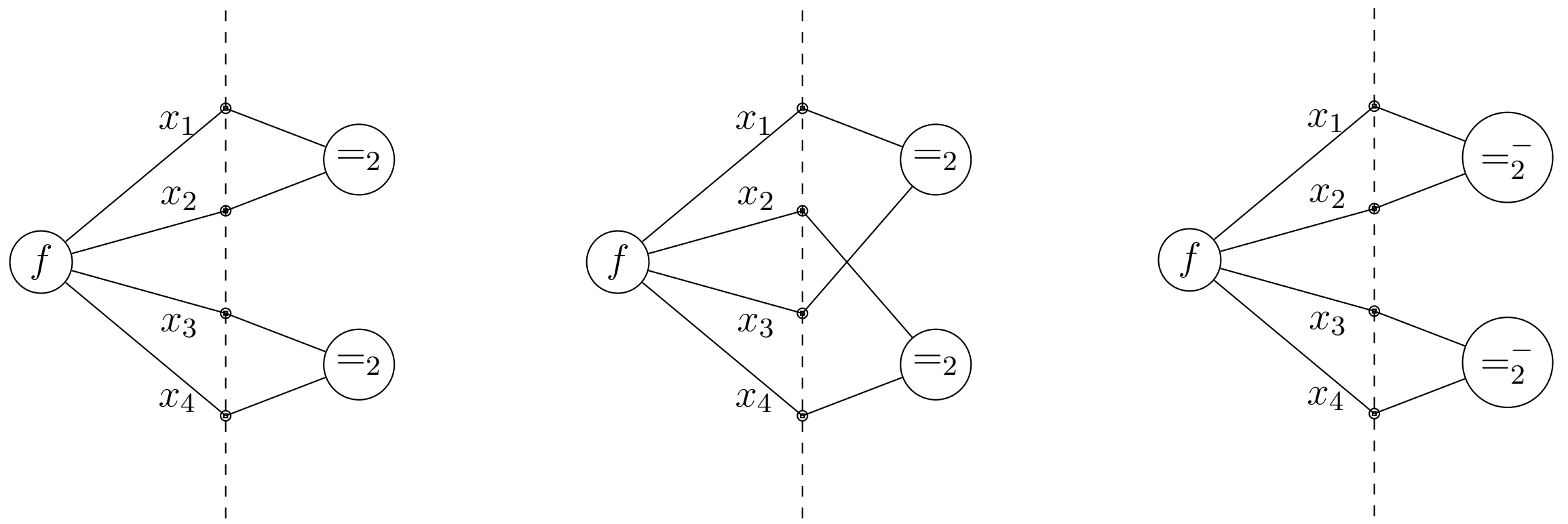
\begin{figure}[H]
\begin{minipage}{0.31\textwidth}
\centering
\begin{tikzpicture}[scale=1]
\node[circle, draw, inner sep=2pt, minimum size=0.6cm] (f) at (0.2,0) {$f$};
\node[circle, draw, inner sep=2pt, minimum size=0.6cm] (g1) at (3.3,1) {$=_2$};
\node[circle, draw, inner sep=2pt, minimum size=0.6cm] (g2) at (3.3,-1) {$=_2$};
\foreach \y/\label in {1.5/x_1, 0.5/x_2, -0.5/x_3, -1.5/x_4} {
    \draw (f) -- (2, \y) 
        node[pos=0.7, \ifdim \y pt > 0 pt above\else below\fi] {$\label$}
        node[draw, inner sep=0.5pt] at (2, \y) {};
}
\draw[dashed] (2,-2.5) -- (2,2.5);
\draw (g1) -- (2,1.5) node[circle,draw,inner sep=1pt]{};
\draw (g1) -- (2,0.5) node[circle,draw,inner sep=1pt]{};
\draw (g2) -- (2,-0.5) node[circle,draw,inner sep=1pt]{};
\draw (g2) -- (2,-1.5) node[circle,draw,inner sep=1pt]{};
\end{tikzpicture}
\textbf{$Z_1=\pij{12}{=_2}\pij{34}{=_2}f$}
\end{minipage}
\quad
\begin{minipage}{0.31\textwidth}
\centering
\begin{tikzpicture}[scale=1]
\node[circle, draw, inner sep=2pt, minimum size=0.6cm] (f) at (0.2,0) {$f$};
\node[circle, draw, inner sep=2pt, minimum size=0.6cm] (g1) at (3.3,1) {$=_2$};
\node[circle, draw, inner sep=2pt, minimum size=0.6cm] (g2) at (3.3,-1) {$=_2$};
\foreach \y/\label in {1.5/x_1, 0.5/x_2, -0.5/x_3, -1.5/x_4} {
    \draw (f) -- (2, \y) 
        node[pos=0.7, \ifdim \y pt > 0 pt above\else below\fi] {$\label$}
        node[draw, inner sep=0.5pt] at (2, \y) {};
}
\draw[dashed] (2,-2.5) -- (2,2.5);
\draw (g1) -- (2,1.5) node[circle,draw,inner sep=1pt]{};
\draw (g1) -- (2,-0.5) node[circle,draw,inner sep=1pt]{};
\draw (g2) -- (2,0.5) node[circle,draw,inner sep=1pt]{};
\draw (g2) -- (2,-1.5) node[circle,draw,inner sep=1pt]{};
\end{tikzpicture}
\textbf{$Z_2=\pij{13}{=_2}\pij{24}{=_2}f$}
\end{minipage}
\quad
\begin{minipage}{0.31\textwidth}
\centering
\begin{tikzpicture}[scale=1]
\node[circle, draw, inner sep=2pt, minimum size=0.6cm] (f) at (0.2,0) {$f$};
\node[circle, draw, inner sep=2pt, minimum size=0.6cm] (g1) at (3.3,1) {$=_2^-$};
\node[circle, draw, inner sep=2pt, minimum size=0.6cm] (g2) at (3.3,-1) {$=_2^-$};
\foreach \y/\label in {1.5/x_1, 0.5/x_2, -0.5/x_3, -1.5/x_4} {
    \draw (f) -- (2, \y) 
        node[pos=0.7, \ifdim \y pt > 0 pt above\else below\fi] {$\label$}
        node[draw, inner sep=0.5pt] at (2, \y) {};
}
\draw[dashed] (2,-2.5) -- (2,2.5);
\draw (g1) -- (2,1.5) node[circle,draw,inner sep=1pt]{};
\draw (g1) -- (2,0.5) node[circle,draw,inner sep=1pt]{};
\draw (g2) -- (2,-0.5) node[circle,draw,inner sep=1pt]{};
\draw (g2) -- (2,-1.5) node[circle,draw,inner sep=1pt]{};
\end{tikzpicture}
$Z_3=\pij{12}{=_2^-}\pij{34}{=_2^-}f$
\end{minipage}
\caption{Three complete contractions of a quaternary signature $f$.
Each network has one $f$ vertex and two binary contraction vertices.
The formulas beneath the diagrams specify the contractions; arity $2d$ uses
$d$ binary contractions.}
\label{fig:expreal}
\end{figure}

We work in the normalized normal basis with available-factor quaternary decompositions, as specified at the start of this section. The first step is the parity constraint.
\begin{lemma}
\label{lem:odd}
Work after the orthogonal normalization, so that the nonzero realizable binary signatures are proportional to $I$ or $\SEq$. Every even-arity signature $f$ in this normalized gadget closure vanishes on odd-weight inputs. The conclusion need not hold before the basis change.
\end{lemma}
\begin{proof}

We induct on the even arity over all signatures in the gadget closure. Let $f$ have arity $2d$.

If $d=1$, $f$ is either $k(=_2)$ or $k(=_2^-)$ with some $k\in\mathbb{C}$, both of which are zero on inputs of odd Hamming weight.

For $d\geq2$, assume the claim holds at arity $2d-2$. Suppose an arity-$2d$ gadget $f$ has $f(\alpha)\ne0$ for an odd-weight string $\alpha$. Since $2d\geq4$, two bits of $\alpha$ agree; after relabeling, take them to be $x_1,x_2$. Contract these ports separately with $=_2$ and $=_2^-$ to obtain two arity-$(2d-2)$ gadgets:
\[
f_0 = \pij{12}{=_2}f, f_1 = \pij{12}{=_2^-}f.
\]
Let $\alpha_0,\alpha_1$ be obtained from $\alpha$ by setting both selected bits to zero or one, respectively, and let $\beta$ be obtained by deleting them. Since those bits agree, $\alpha\in\{\alpha_0,\alpha_1\}$, and $\beta$ still has odd weight. By the definition of contraction,\[
\left\{\begin{aligned}
    f_0(\beta) & = f(\alpha_0) + f(\alpha_1)\\
    f_1(\beta) & = f(\alpha_0) - f(\alpha_1)
\end{aligned}\right.\Rightarrow
\begin{pmatrix}
    f_0(\beta)\\f_1(\beta)
\end{pmatrix} = 
\begin{pmatrix}
    1&1\\1&-1
\end{pmatrix}
\begin{pmatrix}
    f(\alpha_0)\\f(\alpha_1)
\end{pmatrix}
\]
Since $(f(\alpha_0),f(\alpha_1))^T$ is non-zero and $\begin{pmatrix}
    1&1\\1&-1
\end{pmatrix}$ is non-singular, $(f_0(\beta),f_1(\beta))^T$ is also non-zero. So for at least one of $f_0$ and $f_1$, denote it as $g$, there exists $\beta$ such that $|\beta|$ is odd and $g(\beta)\neq0$. $g$ is realized by an actual $\F$-gadget, so $g\in\langle\F\rangle$. That is, we obtain a $2d-2$-ary signature $g$ that violates the induction hypothesis, which leads to a contradiction. Thus, such a signature $f$ does not exist. $f$ is zero on all the inputs of odd Hamming weight.
\end{proof}
By Lemma~\ref{ref:bind}, instances admit a bipartite presentation with binary signatures on one side. Fix an actual signature $f$ of arity $2d$ and close its ports in pairs with $d$ available binaries. Each resulting Holant value is a linear combination of the entries of $f$. Figure~\ref{fig:expreal} illustrates three such complete contractions for a quaternary signature.
The three instances in Figure~\ref{fig:expreal} have values
\[
\left\{
\begin{aligned}
    Z_1&=f_{0000}+f_{0011}+f_{1100}+f_{1111}\\
    Z_2&=f_{0000}+f_{0101}+f_{1010}+f_{1111}\\
    Z_3&=f_{0000}-f_{0011}-f_{1100}+f_{1111}
\end{aligned}\right.\Rightarrow \begin{pmatrix}
    Z_1\\Z_2\\Z_3
\end{pmatrix} = \begin{pmatrix}
    1&1&0&0&1&1\\
    1&0&1&1&0&1\\
    1&-1&0&0&-1&1\\
\end{pmatrix}\begin{pmatrix}
    f_{0000}\\f_{0011}\\f_{0101}\\f_{1010}\\f_{1100}\\f_{1111}
\end{pmatrix},
\]
where $f_\alpha=f(\alpha)$. Only six entries occur in these three tests; the other two even-weight entries are included in the full system below. By Lemma~\ref{lem:odd}, all odd-weight entries vanish. Write $\vec f$ for the vector of even-weight entries and $A$ for the real matrix of all complete $\Eq,\SEq$ contractions. The proof has two steps. First, phase sharing shows that $A\vec f=\lambda\vec v$ for a real vector $\vec v$ and a nonzero scalar $\lambda$ when $f\ne0$. Second, full column rank gives $A\operatorname{Im}(\vec f/\lambda)=0$, hence $f/\lambda$ is real. We establish phase sharing first. For a fixed tensor $f$, let $\mathfrak m$ be a perfect matching of its ports, and let $\ell$ assign a binary contraction label to each pair. Write $Z_f(\mathfrak m,\ell)$ for the resulting complete contraction. Within a pair, order the ports by their indices; reversing that order transposes its binary label.
\begin{lemma}[Sharing Lemma (in normal cases)]
\label{thm:0norm}
Suppose the nonzero binary gadget classes are $[\Eq],[\SEq]$, and every nonzero quaternary gadget is proportional to a tensor product of two members of $\{\Eq,\SEq\}$ in some pairing. For any fixed actual even-arity tensor $f$, any two nonzero values $Z_f(\mathfrak m,\ell)$, with labels in $\{\Eq,\SEq\}$, have a real ratio.
\end{lemma}
The available-factor hypothesis is part of this lemma. Arbitrary quaternary tensor factors are treated in Section~\ref{ssec:normal-enlargement}.
\begin{figure}[t]
\centering
\begin{minipage}{0.30\textwidth}
\centering
\begin{tikzpicture}[scale=1]
\node[circle, draw, inner sep=2pt, minimum size=0.6cm] (A1) at (0.5, 1) {$C_1$};
\node[circle, draw, inner sep=2pt, minimum size=0.6cm] (A2) at (0.5,-1) {$C_2$};
\node[circle, draw, inner sep=2pt, minimum size=0.6cm] (B1) at (3.5, 1) {$B_1$};
\node[circle, draw, inner sep=2pt, minimum size=0.6cm] (B2) at (3.5,-1) {$B_2$};

\coordinate (x1) at (2, 1.5);
\coordinate (x2) at (2, 0.5);
\coordinate (x3) at (2,-0.5);
\coordinate (x4) at (2,-1.5);

\draw (A1) -- (x1) node[pos=0.7, above] {$x_1$} node[draw, inner sep=0.5pt] at (x1) {};
\draw (A1) -- (x2) node[pos=0.7, below] {$x_2$} node[draw, inner sep=0.5pt] at (x2) {};
\draw (A2) -- (x3) node[pos=0.7, above] {$x_3$} node[draw, inner sep=0.5pt] at (x3) {};
\draw (A2) -- (x4) node[pos=0.7, below] {$x_4$} node[draw, inner sep=0.5pt] at (x4) {};

\draw[dashed] (2,-2.5) -- (2,2.5);

\draw (B1) -- (x1) node[circle,draw,inner sep=1pt]{};
\draw (B1) -- (x2) node[circle,draw,inner sep=1pt]{};
\draw (B2) -- (x3) node[circle,draw,inner sep=1pt]{};
\draw (B2) -- (x4) node[circle,draw,inner sep=1pt]{};
\end{tikzpicture}
\textbf{$C_1(x_1,x_2)C_2(x_3,x_4)$}
\end{minipage}
\hfill
\begin{minipage}{0.30\textwidth}
\centering
\begin{tikzpicture}[scale=1]
\node[circle, draw, inner sep=2pt, minimum size=0.6cm] (A1) at (0.5, 1) {$C_1$};
\node[circle, draw, inner sep=2pt, minimum size=0.6cm] (A2) at (0.5,-1) {$C_2$};
\node[circle, draw, inner sep=2pt, minimum size=0.6cm] (B1) at (3.5, 1) {$B_1$};
\node[circle, draw, inner sep=2pt, minimum size=0.6cm] (B2) at (3.5,-1) {$B_2$};

\coordinate (x1) at (2, 1.5);
\coordinate (x2) at (2, 0.5);
\coordinate (x3) at (2,-0.5);
\coordinate (x4) at (2,-1.5);

\draw (A1) -- (x1) node[pos=0.8, above] {$x_1$} node[draw, inner sep=0.5pt] at (x1) {};
\draw (A1) -- (x3) node[pos=0.9, left] {$x_3$} node[draw, inner sep=0.5pt] at (x3) {};
\draw (A2) -- (x2) node[pos=0.9, left] {$x_2$} node[draw, inner sep=0.5pt] at (x2) {};
\draw (A2) -- (x4) node[pos=0.8, below] {$x_4$} node[draw, inner sep=0.5pt] at (x4) {};

\draw[dashed] (2,-2.5) -- (2,2.5);

\draw (B1) -- (x1) node[circle,draw,inner sep=1pt]{};
\draw (B1) -- (x2) node[circle,draw,inner sep=1pt]{};
\draw (B2) -- (x3) node[circle,draw,inner sep=1pt]{};
\draw (B2) -- (x4) node[circle,draw,inner sep=1pt]{};
\end{tikzpicture}
\textbf{$C_1(x_1,x_3)C_2(x_2,x_4)$}
\end{minipage}
\hfill
\begin{minipage}{0.30\textwidth}
\centering
\begin{tikzpicture}[scale=1]
\node[circle, draw, inner sep=2pt, minimum size=0.6cm] (A1) at (0.5, 1) {$C_1$};
\node[circle, draw, inner sep=2pt, minimum size=0.6cm] (A2) at (0.5,-1) {$C_2$};
\node[circle, draw, inner sep=2pt, minimum size=0.6cm] (B1) at (3.5, 1) {$B_1$};
\node[circle, draw, inner sep=2pt, minimum size=0.6cm] (B2) at (3.5,-1) {$B_2$};

\coordinate (x1) at (2, 1.5);
\coordinate (x2) at (2, 0.5);
\coordinate (x3) at (2,-0.5);
\coordinate (x4) at (2,-1.5);

\draw (A1) -- (x1) node[pos=0.8, above] {$x_1$} node[draw, inner sep=0.5pt] at (x1) {};
\draw (A1) -- (x4) node[pos=0.8, left] {$x_4$} node[draw, inner sep=0.5pt] at (x4) {};
\draw (A2) -- (x2) node[pos=0.8, left] {$x_2$} node[draw, inner sep=0.5pt] at (x2) {};
\draw (A2) -- (x3) node[pos=0.8, above] {$x_3$} node[draw, inner sep=0.5pt] at (x3) {};

\draw[dashed] (2,-2.5) -- (2,2.5);

\draw (B1) -- (x1) node[circle,draw,inner sep=1pt]{};
\draw (B1) -- (x2) node[circle,draw,inner sep=1pt]{};
\draw (B2) -- (x3) node[circle,draw,inner sep=1pt]{};
\draw (B2) -- (x4) node[circle,draw,inner sep=1pt]{};
\end{tikzpicture}
\textbf{$C_1(x_1,x_4)C_2(x_2,x_3)$}
\end{minipage}
\caption{Three complete pairings of the open ports of
$g=\mu B_1(x_1,x_2)B_2(x_3,x_4)$. The right vertices are its binary
tensor factors; the left vertices are contraction labels $C_1,C_2$.
These labels may change between the three closures.}
\label{fig:prfreal}
\end{figure}
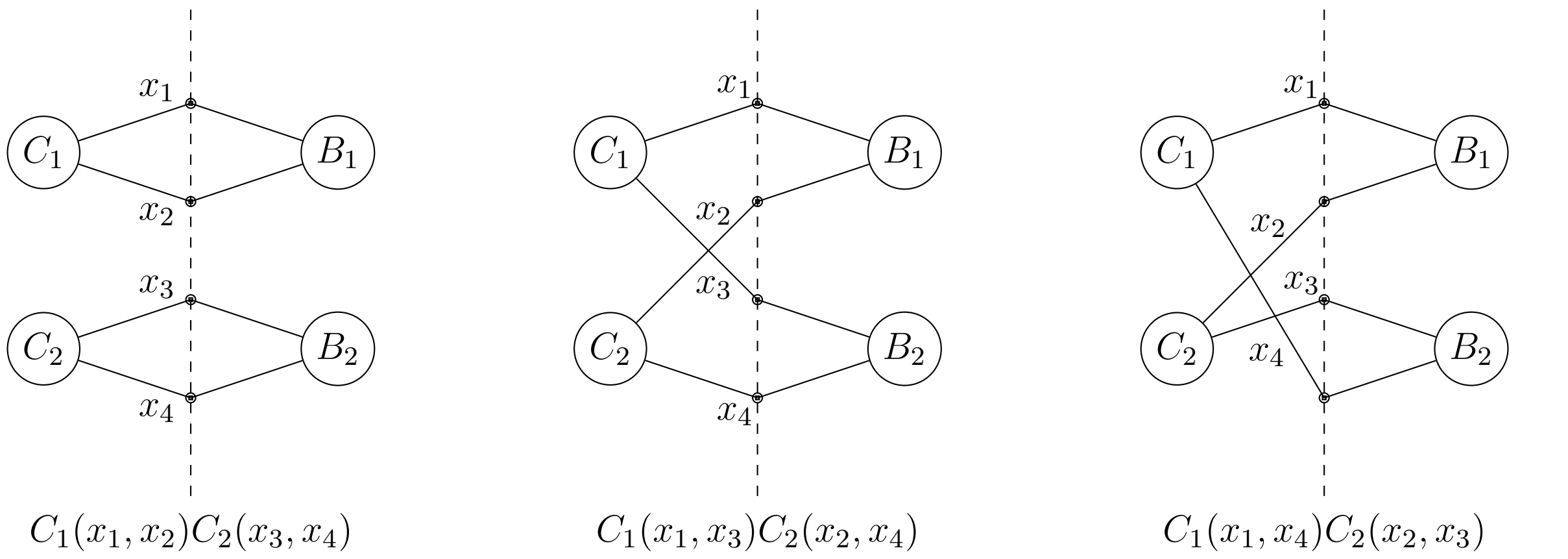
\begin{proof}
We connect the two given nonzero contractions in two stages. At every step the value stays nonzero and changes by a real factor. Figure~\ref{fig:prfreal} shows the possible pairings of the residual factors and contraction labels.

First align their pairings. Removing two contraction edges leaves a nonzero quaternary gadget $g=\mu B_1\otimes B_2$ in some pairing, with $B_1,B_2\in\{\Eq,\SEq\}$. Any desired pairing of these four ports admits a nonzero closure: in the factor pairing use $B_1,B_2$, and in a crossing pairing choose labels whose product is $B_1B_2$. Every such value is a real multiple of $\mu$. Perfect matchings are connected by switches of two edges, so these switches align the pairings while preserving the phase.

Now the two networks have the same pairing. To align their labels, use the formal real contraction $L=I+\SEq$ on at most one special pair. The entrywise contractions satisfy $\langle L,I\rangle =\langle L,\SEq\rangle=2$. Removing a pair with a nonzero closure leaves an actual binary proportional to $I$ or $\SEq$; replacing its label by $L$ therefore preserves nonzeroness and phase. If any labels differ, make one differing pair special and use $L$ there in both networks.

Let $d$ be the number of remaining pairs whose ordinary labels differ. For $d>0$, remove the special pair and one differing ordinary pair from both networks. Their residual quaternary tensors are actual gadgets, since the only formal label has been removed. Their possibilities on these four ports are summarized by
\[
\begin{array}{c|c}
\text{Residual factor pairing} & \text{Nonzero ordinary labels}\\ \hline
\text{Aligned with the two current pairs} & \text{Exactly one of }I,\SEq\\
\text{Crossing the two current pairs} & \text{Both }I,\SEq
\end{array}
\]
An ordinary label working for both residual tensors can therefore be chosen unless both are aligned and require different labels. In that exceptional case, switch the two pairs to a crossing pairing in both networks; now either ordinary label works for both. Use the chosen common ordinary label on one new pair and $L$ on the other. The remaining binary at $L$ is nonzero and proportional to $I$ or $\SEq$, so both new values are nonzero. For either residual tensor, its old and new closures are real multiples of its scalar $\mu$, and hence this step preserves phase.

Both networks still have the same pairing and a common special label. The other pairs are untouched, and $d$ has decreased by one. After finitely many steps the two networks coincide. Their original nonzero values consequently have a real ratio. The formal tensor $L$ is used only as a linear functional; no structural hypothesis has been applied to a gadget containing it.
\end{proof}
Fix a nonzero complete contraction $Z_0$ of $f$, if one exists. The sharing lemma writes every complete contraction as $Z_0$ times a real scalar. Thus $A\vec f=Z_0\vec v$ with $A,\vec v$ real. Here $\vec f$ consists of the even-weight entries, since the other entries vanish by Lemma~\ref{lem:odd}. The next lemma both proves injectivity and guarantees that such a nonzero $Z_0$ exists when $f\ne0$.
\begin{lemma}
In the linear system $\vec Z=A\vec f$, let $\vec f$ contain only the even-weight coordinates, and let $\vec Z$ contain all complete $\Eq,\SEq$ contractions. Then $A$ has full column rank on this even-parity subspace.
    \label{lem:full}
\end{lemma}
\begin{proof}
We prove that the kernel of $A$ on the even-parity subspace is zero. Suppose $A\vec f=0$ and $f\ne0$, where $f$ is any tensor in this subspace. Set $g_0=f$ and choose an even-weight assignment $\alpha^{(0)}$ with $g_0(\alpha^{(0)})\ne0$. At stage $k<d$, maintain a tensor $g_k$ on $2(d-k)$ remaining ports and an even-weight assignment $\alpha^{(k)}$ on those ports such that $g_k(\alpha^{(k)})\ne0$.

There are two remaining ports $i,j$ whose bits in $\alpha^{(k)}$ agree. When at least four ports remain this follows from the pigeonhole principle; when two remain it follows from even parity. Put $\alpha=\alpha^{(k)}$, let $\beta$ be obtained by deleting these two bits, and let $\alpha_{ij}^0,\alpha_{ij}^1$ be obtained by setting both selected bits to zero or one. For $g_k^+=\partial^{\Eq}_{ij}g_k$ and $g_k^-=\partial^{\SEq}_{ij}g_k$,
\[
    \left\{\begin{aligned}
        g_k^+(\beta) & = g_k(\alpha_{ij}^0) + g_k(\alpha_{ij}^1)\\
        g_k^-(\beta) & = g_k(\alpha_{ij}^0) - g_k(\alpha_{ij}^1)
    \end{aligned}\right.\Rightarrow
    \begin{pmatrix}
        g_k^+(\beta)\\g_k^-(\beta)
    \end{pmatrix} = 
    \begin{pmatrix}
        1&1\\1&-1
    \end{pmatrix}
    \begin{pmatrix}
        g_k(\alpha_{ij}^0)\\g_k(\alpha_{ij}^1)
    \end{pmatrix},
    \]
Since $\alpha\in\{\alpha_{ij}^0,\alpha_{ij}^1\}$, the vector on the right is nonzero. The coefficient matrix is nonsingular, so at least one of $g_k^+(\beta),g_k^-(\beta)$ is nonzero. Choose the corresponding contraction as $g_{k+1}$ and set $\alpha^{(k+1)}=\beta$. Deleting two equal bits preserves even weight, so the induction continues.

After $d$ steps, the resulting scalar is a nonzero complete $\Eq,\SEq$ contraction of $f$. This contradicts $A\vec f=0$. Thus $A$ has full column rank. The argument is linear algebra on the whole even-parity subspace and does not assume that the tensor being tested is realizable.
\end{proof}

For nonzero $f$, choose a nonzero complete contraction $Z_0$. Since $A(\vec f/Z_0)$ is real and $A$ is real with full column rank, $A\operatorname{Im}(\vec f/Z_0)=0$ implies that $f/Z_0$ is real. This proves the claimed realnumberization without changing the signature's basis or adjoining any formal linear combination.

\begin{theorem}[Normal branch with available binary factors]
\label{thm:de_normal}
Assume the nonzero binary classes in the normalized normal basis are $[\Eq],[\SEq]$, and every nonzero realizable quaternary signature is a scalar multiple of a product of two available binary signatures in some variable pairing. Then $\Holant(\F)$ is in $\mathrm{FP}$ or is \#P-hard.
\end{theorem}
\begin{proof}
The sharing lemma and the full-column-rank lemma make each nonzero input signature real after division by a nonzero complete contraction. Apply Lemma~\ref{lem:real} to the entire finite signature set.
\end{proof}

Theorem~\ref{thm:de_normal} treats quaternary gadgets whose binary tensor factors are available. To handle arbitrary factors, we introduce the Pauli-basis tools used in both the normal and isotropic cases.

\subsection{Pauli Basis and other tools}
\label{ssec:pauli-tools}
\label{ssec:normal-enlargement}
Keep $\Pa=\{I,X,Y,Z\}$ as in Section~\ref{sec:prel}, and denote its real scalar-normalized representatives by $\mathcal B$:
\[
\mathcal B=\{\Eq,\SEq,\NEq,\SNEq\},\qquad
M(\mathcal B)=\{I,-iZ,-iX,Y\}.
\]
The Pauli matrices in $\Pa$ form a basis of the space of complex binary signature matrices, while their scalar-normalized representatives in $\mathcal B$ form a basis of the space of real binary tensors. For any fixed pairing of $2d$ ports, the tensor products of $d$ members of $\mathcal B$ therefore form a basis of the space of real $2d$-ary tensors. This gives the full-rank reconstruction used in Corollary~\ref{cor:iso-pauli-real}. These algebraic basis elements are not assumed to be available gadgets; availability is checked in each application.

A signature of arity $2d$ can be represented as a $2^d\times 2^d$ matrix by partitioning its variables into two ordered groups of size $d$. These matrices form a complex vector space of dimension $4^d$. For a fixed pairing of the ports, write $f=\sum_\alpha c_\alpha P_\alpha$, where each $P_\alpha$ is a tensor product of $d$ members of $\Pa$. A signature grid $\Omega$ with label $f$ at vertex $u$ then satisfies
\[
Z(\Omega)=\sum_\alpha c_\alpha
\sum_{\sigma:E\to\{0,1\}} P_\alpha(\sigma|_{E(u)})
\prod_{v\ne u} f_v(\sigma|_{E(v)}).
\]
This is linearity at one vertex, with $P_\alpha$ evaluated on the ports of that vertex. Expanding every vertex can produce exponentially many terms; here the basis is used for linear reconstruction. Replacing $X,Z$ by $-iX,-iZ$ gives the real basis $\mathcal B$; the corresponding powers of $i$ are absorbed into the coefficients.

The tensor-power lemma~\cite[Lemma 33]{meng2025odd} gives, without the one-sided exception of Lemma~\ref{lem:gendecomp},
\[
\Holant(\G,u)\le_T\Holant(\G,u^{\otimes r})\quad(r\ge1).
\]
The same reduction holds with $\NEq$ edges, by the holographic transformation $K$ introduced in Section~\ref{sssec:holographic-transformations}.

We use it for binary tensor squares, and prove the enlarged gadget structure separately. Odd factors and one-sided tails in the isotropic case will instead be handled by an actual binary obstruction and a direct weight-counting argument. A common orthogonal transformation making every signature real up to its own nonzero scalar invokes the real Holant dichotomy, as in Lemma~\ref{lem:real}. We prove the required realnumberization; the presence of an auxiliary binary alone is not used as a complexity classification.

The next sharing lemma uses ordinary equality edges and the pairing and label notation of Lemma~\ref{thm:0norm}. It applies to the isotropic binary classes in the original basis and to the real four-class extension.
\begin{lemma}[Sharing Lemma in isotropic case]\label{lem:iso-sharing}
Let $\mathcal B$ be either $\{\Eq,\SNEq\}$ or $\{\Eq,\SEq,\NEq,\SNEq\}$. Assume every member of $\mathcal B$ is available up to a nonzero scalar. Suppose every nonzero binary gadget is proportional to a member of $\mathcal B$, and every nonzero quaternary gadget is proportional to a product of two members of $\mathcal B$ in some pairing. For any fixed actual even-arity tensor $v$, all its nonzero complete $\mathcal B$ contractions $Z_v(\mathfrak m,\ell)$ have real ratios.
\end{lemma}
\begin{proof}
We again align pairings first and labels second, preserving a nonzero value up to a real factor at every step.

Both sets consist of real non-singular matrices, are closed under multiplication and transposition up to a real sign, and are mutually orthogonal for the entrywise pairing, with squared norm two. Two elementary consequences will be used. A binary factor is closed by exactly its own label. A product of two factors, closed using a crossing pairing, leaves a non-singular member of $\mathcal B$ after either crossing pair receives any label in $\mathcal B$. Thus any pairing of its four ports admits a nonzero closure. Every such closure is a real multiple of the product's common scalar.

Take two nonzero complete contractions of $v$. First align their pairings by successive two-edge switches. Remove the two edges of a switch. The remaining nonzero quaternary gadget has the stated product form, so labels for the desired pairing can be chosen to keep the value nonzero. Each switch preserves its complex phase up to a real factor. Perfect matchings are connected by these switches, so the two contractions now have the same pairing.

It remains to align labels without passing through zero. Set $L=\sum_{B\in\mathcal B}B$, and permit this formal real tensor on one special pair only. Orthogonality gives $\langle L,B\rangle=2$ for every $B\in\mathcal B$. If the labels differ, choose a differing pair as special. Removing it from either network leaves an actual binary gadget. Replacing its label by the same $L$ in both networks keeps both values nonzero and preserves their phases.

Remove this special pair and another pair whose ordinary labels differ. The two residual quaternary gadgets are actual, since the only formal label has been removed. In a crossing factorization every ordinary label on the nonspecial pair leaves a nonzero binary on the special pair; in an aligned factorization exactly one label does. A common ordinary label therefore works for both gadgets unless both factorizations are aligned and require different labels. In that exceptional case switch the two pairs to a crossing pairing for both gadgets; every ordinary label then works for both. Use $L$ on the new special pair; its pairing with either remaining binary factor is nonzero. All old and new values are real multiples of the respective quaternary gadget's scalar, so both phases are preserved. Let $d$ count ordinary pairs with differing labels. The operation decreases $d$ by one, while retaining a common pairing and just one common special pair. Every other pair is untouched, so repetition terminates with $d=0$ and makes the networks identical, proving that their original values have a real ratio. No structural hypothesis was applied to a gadget containing the formal $L$ label.
\end{proof}
Taking all four real contraction forms gives the following consequence.
\begin{corollary}\label{cor:iso-pauli-real}
Under the four-matrix hypotheses of Lemma~\ref{lem:iso-sharing}, every signature is real up to a nonzero scalar. The corresponding Holant problem is therefore in FP or is \#P-hard.
\end{corollary}
\begin{proof}
The four real matrices $\Eq,\SEq,\NEq,\SNEq$ form a basis of all real binary tensors. On any fixed pairing, their tensor products span the entire real space of even-arity tensors. Thus the matrix $A$ of these complete contractions has full column rank over $\mathbb R$. For a nonzero input $f$, at least one contraction is nonzero; Lemma~\ref{lem:iso-sharing} gives $Af=\lambda r$ with $r$ real. Write $f/\lambda=f_R+ig$, with $f_R,g$ real. Since $A$ is real, $Ag=0$, so full rank gives $g=0$. Apply the real Holant dichotomy to the entire finite signature set, with its individual scalar normalizations.
\end{proof}

The following observation excludes unary factors by an actual gadget construction. It will let us complete the normal case without invoking the one-sided exception of the generalized decomposition theorem.
\begin{lemma}\label{lem:unary-factor-obstruction}
Suppose $\Eq,\SEq$ are available and every nonzero binary gadget is nonsingular. No nonzero quaternary gadget has a unary tensor factor. Consequently every nonzero decomposable quaternary gadget is a product of two nonsingular binary tensors.
\end{lemma}
\begin{proof}
Suppose a quaternary gadget factors as $u\otimes h$, where $u$ is a nonzero unary and $h$ is a nonzero ternary tensor. Choose $x\in\operatorname{supp}(h)$. Two of its three bits agree; call their positions $i,j$ and the remaining bit $b$. Let $h_{00b},h_{11b}$ denote the entries obtained by setting the selected two bits to $00,11$, respectively, while retaining $b$ at the third position. At least one of these entries is nonzero. The corresponding entries of $\partial^{\Eq}_{ij}h$ and $\partial^{\SEq}_{ij}h$ are $h_{00b}+h_{11b}$ and $h_{00b}-h_{11b}$, so at least one of these unary contractions, say $v$, is nonzero. Performing that contraction on the actual quaternary gadget realizes $u\otimes v$, a nonzero rank-one binary. This contradicts the stated binary hypothesis. Neither $u$ nor $h$ has been adjoined as a separate signature.

A nontrivial partition of four ports has sizes $1+3$ or $2+2$. The first is excluded. A singular nonzero binary factor in the second would split into unaries and again give a unary factor of the quaternary tensor. Thus both binary factors are non-singular.
\end{proof}

We can now extend the available-factor normal proof to arbitrary tensor decomposition.
\begin{proposition}\label{prop:normal-enlargement}
Suppose the binary classes are $[\Eq],[\SEq]$ and every quaternary gadget is tensor-decomposable. Either realnumberization gives an FP versus \#P-hardness classification of $\Holant(\F)$, or $\Holant(\F)\equiv_T\Holant(\mathcal G)$ for an enlarged signature set $\mathcal G$ whose binary classes are
\[
[\Eq],\ [\SEq],\ [A],\ [\SEq A],\qquad
A=\begin{pmatrix}0&1\\i&0\end{pmatrix},
\]
and every nonzero quaternary gadget is a product of two members of these four classes, up to scalar and variable pairing.
\end{proposition}
\begin{proof}
By Lemma~\ref{lem:unary-factor-obstruction}, each nonzero decomposable quaternary gadget is a product of two nonsingular binaries. Thus no odd-arity factor or one-sided decomposition exception occurs here. The even-parity constraint of Lemma~\ref{lem:odd} forces both factors to have fixed, equal parity. Diagonal factors can be isolated by a nonzero $I$ or $\SEq$ contraction of the other factor, so they already belong to the available classes. If every factor is diagonal, Theorem~\ref{thm:de_normal} applies.

For an antidiagonal pair, write its entries as $B=(0,a,b,0)$ and $C=(0,c,d,0)$, with $abcd\ne0$. The two crossing $I$ contractions have diagonal entries $(ac,bd)$ and $(ad,bc)$. Hence $ac=\sigma bd$ and $ad=\tau bc$ for $\sigma,\tau\in\{1,-1\}$, and $a^2=\sigma\tau b^2$. Rescale $B$ to $A_\rho=(0,1,\rho,0)$, where $\rho\in\{1,-1,i,-i\}$. The relation $d/c=\tau b/a$ makes $C$ proportional to $B$ or to $B$ with a local $\SEq$. We therefore have an actual $A_\rho\otimes A_\rho$, and the tensor-power lemma adjoins $A_\rho$.

We check the enlarged gadget closure without assuming that adjoining a factor preserves $C_2$. An even number of new $A_\rho$ vertices can be replaced in pairs by the original tensor-square gadget. For an odd number in a gadget with nonempty boundary, attach one extra $A_\rho$ to a boundary port and then perform these replacements. This is an invertible local operation, since $A_\rho^2=\rho I$. It follows that all enlarged quaternary gadgets remain decomposable, and that their binary classes are exactly $[\Eq],[\SEq],[A_\rho],[\SEq A_\rho]$. These four matrices form a basis of binary tensors. A unary factor $u$ of an enlarged quaternary gadget would leave a nonzero ternary complement. Some available binary contraction of two ports of that complement is a nonzero unary $v$, giving the forbidden rank-one binary $u\otimes v$. Thus a nontrivial quaternary factorization consists of two binaries. Some available contraction of either factor isolates the other, so both belong to the four nonsingular enlarged binary classes.

For $\rho=\pm1$, these are the real Pauli classes, and Corollary~\ref{cor:iso-pauli-real} classifies the corresponding Holant problem. For $\rho=\pm i$, multiplying by $\SEq$ interchanges the two choices, giving the displayed representative $A$.
\end{proof}

The enlarged signature set can be taken to be $\mathcal G=\F\cup\{A\}$, with the already realizable $\Eq,\SEq$ included. Its four-class closure was proved by the tensor-square replacement argument; it was not inferred from Turing equivalence. In this extension the available contraction forms are $\Eq,\SEq,A,\SEq A$. In particular, $\SNEq$ is not available. Although
\[
\SNEq=\frac{1+i}{2}A+\frac{1-i}{2}\SEq A,
\]
this linear identity does not realize its left-hand side as a gadget: gadget closure does not imply closure under addition. Thus the four-class assertion concerns contractions by its four available classes, not by arbitrary members of the Pauli basis. For example, the formal contraction $\partial^{\SNEq}_{14}(A_{12}A_{34})=\NEq_{23}$ lies outside these four classes; it is not an available operation in this extension.

Every member of $\F$ has even parity, whereas $A$ has odd parity. Consequently every gadget over $\mathcal G$ has a definite parity: the parity of its external assignment is the sum of its vertex parities, since every internal edge contributes twice. We now classify $\Holant(\mathcal G)$ directly. No enlargement obtained merely by a complexity reduction is used as a contradiction to the original $C_2$ hypothesis.

Put $\mathcal B_A=\{\Eq,\SEq,A,\SEq A\}$, and let $\mathcal M^A_n$ contain zero and all scalar multiples of products of $n/2$ members of $\mathcal B_A$ on disjoint variable pairs. The four matrices form a basis of binary tensors. They are nonsingular and closed under multiplication, inversion and transposition up to nonzero scalar. Indeed, $(\SEq)^2=I$, $A^2=iI$, $A\SEq =-\SEq A$ and $A^T=i\SEq A$. In particular, local attachments from $\mathcal B_A$ preserve $\mathcal M^A_n$ and preserve the condition that all single-pair $\mathcal B_A$ contractions lie in $\mathcal M^A_{n-2}$. The following argument adapts the factor-integration method of~\cite{cai2020eoars}; here the full binary basis makes it possible to reconstruct a common factor directly from four slices.

\begin{lemma}[Factor propagation above arity eight]
\label{lem:normal-high}
Let $n\ge10$ be even. If every single-pair $\mathcal B_A$ contraction of a tensor $f$ belongs to $\mathcal M^A_{n-2}$, then $f\in\mathcal M^A_n$.
\end{lemma}
\begin{proof}
Assume $f\ne0$. Since $\mathcal B_A$ spans all binary tensors, some contraction is nonzero. Local attachments, a port permutation and a scalar normalization put it in the form
\[
\partial^{\Eq}_{12}f=(\Eq)_{34}(\Eq)_{56}(\Eq)_{78}(\Eq)_{9,10}\cdots (\Eq)_{n-1,n}.
\]
Write $\partial_{ij}=\partial^{\Eq}_{ij}$ in this proof.

We use the following elementary propagation rule. Let $R$ be a nonzero product of nonsingular binary factors. Suppose contractions on two different pairs of $R$, neither deleting $u,v$, are nonzero and both have a factor $C_{uv}$. Then $C_{uv}$ is a factor of $R$. Indeed, if $u,v$ belonged to different factors of $R$, a contraction could join them into one factor only by deleting their two partners. That uniquely determines the deleted pair. If they are already paired in $R$, their factor survives either contraction and is proportional to $C$. Here we use uniqueness of the prime tensor factorization~\cite{meng2025odd}; a nonsingular binary is tensor-prime.

Set $R=\partial_{34}f$. It is nonzero and is a binary product, because
\[
\partial_{12}R=2(\Eq)_{56}(\Eq)_{78}(\Eq)_{9,10}\cdots (\Eq)_{n-1,n}.
\]
If $1,2$ are paired in $R$, all its other pairs are displayed in this equation. Otherwise their partners form exactly one displayed pair, which we may name $5,6$. In either case $(\Eq)_{78}$ and $(\Eq)_{9,10}$ divide $R$, and $6$ is paired with neither $7$ nor $8$. Thus the $67,68,78$ contractions of both $R$ and $\partial_{12}f$ are nonzero and retain $(\Eq)_{9,10}$. Commuting contractions and applying the propagation rule gives
\[
(\Eq)_{9,10}\mid\partial_{67}f,\qquad
(\Eq)_{9,10}\mid\partial_{68}f,\qquad
(\Eq)_{9,10}\mid\partial_{78}f.
\]

For any $B\in\mathcal B_A$, consider $S=\partial^B_{12}f$. If $S=0$, it is divisible by $(\Eq)_{9,10}$. Otherwise $S$ is a product of nonsingular binaries. A contraction of two variables from different factors is nonzero. Among the three overlapping pairs $67,68,78$, at most one can be a factor pair of $S$; hence at least two of its corresponding $I$ contractions are nonzero. By the displayed divisibilities and commutation, both retain $(\Eq)_{9,10}$. The propagation rule implies $(\Eq)_{9,10}\mid S$.

This holds for all four $B$. Since their contraction functionals span the dual binary space on ports $1,2$, reconstructing $f$ from these four slices gives $(\Eq)_{9,10}\mid f$. Contracting that factor by $I$ isolates twice its complement, which belongs to $\mathcal M^A_{n-2}$ by hypothesis. Therefore $f\in\mathcal M^A_n$.
\end{proof}

\begin{lemma}[The two nonstandard finite cases]
\label{lem:normal-finite}
Let $n\in\{6,8\}$. Suppose $f$ has a definite parity and every single-pair $\mathcal B_A$ contraction belongs to $\mathcal M^A_{n-2}$. Then $f\in\mathcal M^A_n$.
\end{lemma}
\begin{proof}
The zero tensor is immediate. As in Lemma~\ref{lem:normal-high}, normalize a nonzero contraction to
\[
\partial^{\Eq}_{12}f=P:=(\Eq)_{34}(\Eq)_{56}\cdots(\Eq)_{n-1,n}.
\]
Local attachments preserve definite parity, and this nonzero contraction makes the normalized tensor even. Its coordinates are therefore indexed by the $2^{n-1}$ even-weight words.

We describe explicitly the finite alternatives for a contraction. For a nonzero product $R=\lambda\prod_{uv} C_{uv}$, let $\mathscr L(R)$ be the following set of products on two additional ports $1,2$:
\[
\frac12(\Eq)_{12}R,\qquad
\lambda B_{1u}(B^{-T}C)_{2v}\prod_{ab\ne uv}C_{ab}.
\]
In the second expression choose a factor $C_{uv}$ of $R$, either order of its two ports, and $B\in\mathcal B_A$. Matrices are identified up to scalar when choosing their class; the displayed expression retains the scalar. Contraction at $1,2$ gives $R$, since $B^T(B^{-T}C)=C$. Conversely, in a binary product whose $1,2$ contraction equals $R$, either $1,2$ form one factor, or their two partners form one factor of $R$. These are precisely the displayed alternatives. Thus $\mathscr L(R)$ is the complete set of product preimages of $R$, with $1+4\,\operatorname{arity}(R)$ members.

For $ij\subseteq\{3,\ldots,n\}$, commuting contractions gives
\[
\partial^{\Eq}_{ij}f\in
\mathscr L(\partial^{\Eq}_{ij}P),\qquad
\partial^{\SEq}_{35}f\in
\mathscr L(\partial^{\SEq}_{35}P).
\]
Every argument of $\mathscr L$ here is nonzero. The lists have nine members for $n=6$ and seventeen for $n=8$. These are finite affine equations, not restrictions on possible coefficient fields of $f$. To specify their coefficient matrices, for an even word $x$ and a word $y$ on the remaining ports put
\[
D_{ij}^{B}[y,x]=B(x_i,x_j)\,
\mathbf1_{x|_{\overline{ij}}=y}.
\]
The normalization is $D_{12}^{\Eq}f=P$. Successive substitution of the displayed lists for $D_{ij}^{\Eq}f$, with the disjoint pairs in lexicographic order, followed by $D_{35}^{\SEq}f$, gives the following affine systems:
\[
\begin{array}{c|c|cc|cc|c}
n&2^{n-1}&\multicolumn{2}{c|}{\text{disjoint pairs}}
 &\multicolumn{2}{c|}{\text{signed pair }35}
 &D_{12}^{A}f\in\mathcal M^A_{n-2}\\
 &&\text{rank}&\text{systems}&\text{rank}&\text{points}
 &\text{points}\\ \hline
6&32&29&17&32&33&17\\
8&128&125&33&128&33&25
\end{array}
\]
Here identical affine spaces are counted once. The substitution uses only the entries $0,1,-1,i,-i$ in the four binary matrices: for each candidate right-hand side, eliminate a coordinate with nonzero coefficient and substitute it into the remaining equations; a zero row requires a zero right-hand side. All final consistent systems have full column rank. Imposing the last column leaves exactly
\[
f\in\mathscr L(P).
\]
In particular, these points are explicitly the paired product $(\Eq)_{12}P/2$ and the crossed products in the definition of $\mathscr L(P)$, one for each factor of $P$, orientation and binary label. Their number is $1+8(n/2-1)=4n-7$. Every such product satisfies all the remaining contraction conditions by closure of $\mathcal B_A$. Since all the equations are over $\mathbb Q(i)$ and have full column rank, the same classification holds over $\mathbb C$. It follows that $f\in\mathcal M^A_n$.
\end{proof}

\begin{theorem}[Polynomial-time evaluation of the nonstandard extension]
\label{thm:normal-four-class}
Let $\mathcal G$ be a finite set of even-arity signatures, each with definite parity. Suppose its nonzero binary gadget classes are $\mathcal B_A$, and all its quaternary gadgets belong to $\mathcal M^A_4$. Then every input signature belongs to $\mathcal M^A$, and $\Holant(\mathcal G)$ is in FP.
\end{theorem}
\begin{proof}
Each gadget has definite parity, by summing its vertex parities. If a nonzero gadget outside $\mathcal M^A$ existed, choose one of minimum arity $n$. The binary and quaternary hypotheses give $n\ge6$. Every single-pair $\mathcal B_A$ contraction has smaller arity and lies in $\mathcal M^A_{n-2}$. Lemma~\ref{lem:normal-finite} handles $n=6,8$, and Lemma~\ref{lem:normal-high} handles $n\ge10$. Both contradict its choice. Thus all inputs are binary products.

Factor each fixed input signature once. Splitting its occurrences into binary vertices reduces an instance to cycles. Each cycle is evaluated by the trace of a product of $2\times2$ matrices, with transposes recording port orientation. The known scalar factors are multiplied back in. This takes polynomial time and does not require the individual tensor factors to be original gadgets.
\end{proof}

\begin{corollary}[The normal decomposable dichotomy]
\label{thm:normal-decomp-all}
Let $\F$ be a finite even-arity signature set whose nonzero binary gadget classes are $[\Eq],[\SEq]$. If every quaternary gadget is tensor-decomposable, then $\Holant(\F)$ is in $\mathrm{FP}$ or is \#P-hard.
\end{corollary}
\begin{proof}
Proposition~\ref{prop:normal-enlargement} either gives the FP versus \#P-hardness classification by realnumberization or gives the nonstandard extension. Its inputs have definite parity, as proved above. Theorem~\ref{thm:normal-four-class} evaluates that extension in FP, so $\Holant(\F)$ is also in FP. This uses a classification of the enlarged problem, not a contradiction between its four binary classes and the original two-class hypothesis.
\end{proof}

\subsection{Isotropic Case}
\label{ssec:isotropic}
We continue the decomposable-quaternary case of Section~\ref{sec:4}. Decomposition means arbitrary tensor decomposition; a factor is not assumed to be an available gadget. Instances containing a zero signature have value zero; nullary signatures contribute separately computable scalar factors.

\begin{theorem}[Decomposable isotropic dichotomy]
\label{thm:iso-complete}
Suppose the nonzero binary classes in the gadget closure of an even-arity signature set $\F$ are $[\Eq],[\SNEq]$, where $\SNEq=\left(\begin{smallmatrix}0&1\\-1&0\end{smallmatrix}\right)$. If every nonzero quaternary gadget is tensor-decomposable, then $\Holant(\F)$ is in $\mathrm{FP}$ or is \#P-hard.
\end{theorem}

We first classify the possible quaternary tensor factors. In the remaining branch, a minimum-arity gadget outside the binary matching products has either a product middle layer or a Reed--Muller middle layer. The product case forces a common one-sided support condition; the Reed--Muller case supplies additional contraction tests. In both cases phase sharing and injectivity make the relevant signatures real up to scalars, allowing the real Holant dichotomy to apply.

Use the matrix $K$ and the notation $\hat f=(K^{-1})^{\otimes n}f$ for an arity-$n$ signature from Section~\ref{sssec:holographic-transformations}. The identities needed for the isotropic case are
\[
\Holant(\F)\equiv_T\Holant(\NEq\mid\hat\F),\qquad
K^TK=\NEq,\quad K^{-1}I(K^{-1})^T=\NEq,\quad
K^{-1}\SNEq(K^{-1})^T=i\SNEq.
\]

All subsequent signatures, unless specified otherwise, are on the right of this bipartite presentation. Binary composition is $B\NEq C$. Connecting a right-hand binary $B$ to two ports induces the contraction form $\NEq B\NEq $. Since $\NEq \NEq \NEq =\NEq$ and $\NEq \SNEq \NEq =-\SNEq$, both $\NEq$ and $\SNEq$ contractions are available up to a tracked scalar. Attaching $\SNEq$ to one port acts as $\SEq$ up to sign, so independent local $\SEq$ transformations and their inverses are actual gadgets. We normalize nonzero constant scalars freely, always retaining them in reductions.

Let $\mathcal M_n$ consist of zero and all scalar multiples of products of $n/2$ members of $\{\NEq,\SNEq\}$ on disjoint pairs, in any pairing. Put $\mathcal M=\bigcup_{n\text{ even}}\mathcal M_n$, with scalars included at arity zero. Products and local $\SEq$ transformations preserve this class. Nonzero members have a unique pairing: their nonsingular binary factors are tensor-prime. An $\NEq$ contraction of such a product is zero exactly when the contracted ports form one of its $\SNEq$ factors. A contraction between different factors joins two nonsingular antidiagonal matrices and is nonzero.

For $n=2d$, write $h_{=}$ for the restriction of $h$ to weight $d$, extended by zero. Complete $\NEq,\SNEq$ contractions depend only on $h_{=}$ and determine it uniquely. Indeed, for any weight-$d$ string, pair its zeros with its ones; the tensor product of the oriented tests $(\NEq+\SNEq)/2,(\NEq-\SNEq)/2$ extracts that coordinate. Thus their joint map has rank $\binom{2d}{d}$ and kernel exactly the noncentral layers. These tests are linear functionals. The structural argument below determines when the unseen layers can affect a closed Holant value.

\begin{lemma}\label{lem:iso-parity4}
Every realizable transformed quaternary signature has even parity.
\end{lemma}
\begin{proof}
In an odd-weight four-bit string choose a mixed pair. Its $\NEq,\SNEq$ contractions at the remaining two equal bits give $u+v,u-v$. Both are zero because each realizable binary is zero or proportional to $\NEq,\SNEq$. Therefore $u=v=0$.
\end{proof}

\begin{lemma}\label{lem:iso-diagonal-factor}
Assume every nonzero quaternary gadget is tensor-decomposable. Either the corresponding Holant problem is in FP or is \#P-hard, or every nonzero transformed quaternary gadget belongs to $\mathcal M_4$ or is a scalar multiple of $\delta_{0^4}$ or $\delta_{1^4}$.
\end{lemma}
\begin{proof}
Suppose first that a nonzero quaternary gadget has a unary factor, and write it as $u\otimes v$ with $v$ ternary. Every $\NEq$ or $\SNEq$ contraction within $v$ must vanish: otherwise the same actual contraction of the quaternary gadget is a nonzero rank-one binary. The two contraction forms separate the entries $01$ and $10$ on each pair. Hence $v$ is supported on $000,111$ only. Write $u=(s,t)$ and $v=a\delta_{000}+b\delta_{111}$. Even parity of $u\otimes v$ gives $ta=sb=0$. At most one of $sa,tb$ is nonzero, so the quaternary gadget is a pure extreme power as claimed. This argument never adjoins either factor.

Otherwise its prime factors are two nonsingular binaries: any singular binary factor would have a unary factor. Parity forces each factor to have fixed parity, and their parities agree. If both factors are antidiagonal, some $\NEq$ or $\SNEq$ contraction of either factor is nonzero and isolates the other as an actual binary. Thus both belong to the current binary classes, giving $\mathcal M_4$.

In the remaining case write the gadget as $D_1\otimes D_2$, where $D_1=\diag(a,b)$, $D_2=\diag(c,d)$ and $abcd\ne0$. A crossing $\NEq$ contraction has antidiagonal entries $ad,bc$, so $ad=\pm bc$. Consequently $D_2$ is proportional to $D_1$ or $D_1\SEq$. A local $\SEq$ gives an actual $D\otimes D$, with $D=\diag(a,b)$. For this square, the contractions are explicitly
\[
\partial^{\NEq}_{12}(D_{12}D_{34})
=\partial^{\SNEq}_{12}(D_{12}D_{34})=0,\qquad
\partial^{\NEq}_{13}(D_{12}D_{34})=ab\,\NEq_{24},\qquad
\partial^{\SNEq}_{13}(D_{12}D_{34})=ab\,\SNEq_{24}.
\]
The other factor contraction also vanishes, and the other crossing contractions give the same two classes. These are contraction forms; realizing the $\SNEq$ form by a right-hand $\SNEq$ vertex introduces the known minus sign from $\NEq\SNEq\NEq=-\SNEq$. Apply $S=\diag(a^{-1/2},b^{-1/2})$ to all right-hand signatures. The new left kernel is $S^{-T}\NEq S^{-1}=\sqrt{ab}\,\NEq$, whose known edge scalar is harmless, and $D\otimes D$ becomes $I\otimes I$. The tensor-power lemma adjoins $I$. Denote the normalized signature set before this adjoining step by $\mathcal H$.

We check the enlarged signature set's structure. In a gadget over $\mathcal H\cup\{I\}$ with an even number of new $I$ vertices, replace pairs of them by the actual $I\otimes I$ gadget. If their number is odd and the boundary is nonempty, attach one more $I$ through an $\NEq$ edge to a boundary port and then perform the replacements. That attachment acts as the invertible matrix $\NEq$ on that port. Hence every enlarged quaternary gadget remains decomposable. The same argument on binary gadgets restricts their classes to $[\Eq],[\SEq],[\NEq],[\SNEq]$. All four occur, since the first two original classes are $[\NEq],[\SNEq]$, $I$ was adjoined, and $I\NEq \SNEq =-\SEq$.

Convert to ordinary equality edges: an $I$ vertex between two $\NEq$ edges supplies an equality wire, and conversely an $\NEq$ edge can be represented by an ordinary $\NEq$ vertex. Binary and quaternary boundary tensors are preserved. The enlarged ordinary signature set has $\Eq,\SEq$ available and only nonsingular nonzero binaries, so Lemma~\ref{lem:unary-factor-obstruction} excludes unary factors. Every nonzero quaternary gadget is therefore a product of two nonsingular binaries. The four available matrices span all binary tensors, so some available contraction of either factor is nonzero and isolates the other. Its factors therefore belong to the four available classes. Corollary~\ref{cor:iso-pauli-real} now applies to this enlarged ordinary signature set and completes the classification.
\end{proof}

We next exclude opposite weight directions without adjoining a pinning signature. This is the step that makes a direct FP upper bound possible for the one-sided branch.

\begin{lemma}\label{lem:iso-tail-obstruction}
Suppose the nonzero binary gadget classes, with $\NEq$ edges, are $[\NEq],[\SNEq]$. Let $p(t)$ be a tensor polynomial of even arity $m\ge2$ such that $p(t)$ is an actual gadget, up to a known nonzero scalar, for every positive integer $t$. If its leading coefficient is $a\delta_{0^m}$ with $a\ne0$, every signature in the gadget closure has support below or at half weight. The analogous statement with $\delta_{1^m}$ gives support above or at half weight. A constant polynomial includes an actual pure pinning power.
\end{lemma}
\begin{proof}
Suppose an actual arity-$2d$ tensor has a nonzero entry of weight $d+k$, where $k>0$. Pair each of its $d-k$ zero positions with a one position. The oriented coordinate tests $(\NEq+\SNEq)/2,(\NEq-\SNEq)/2$ on these pairs isolate that entry at the remaining $2k$ ones. Expanding the tests as a linear combination of actual $\NEq,\SNEq$ contractions shows that at least one actual tensor $q$ of arity $\ell=2k$ satisfies $q(1^\ell)\ne0$.

Temporarily use $\delta_{0^m}$ as a formal label. Let $c=\gcd(m,\ell)$, take $s=m/c$ copies of $q$ and $r=\ell/c$ copies of this formal label, and match all $s\ell=rm$ ports of the first group to those of the second through $\NEq$ edges. Cut one such edge to obtain a binary tensor $B$; order its open ports as the $q$ port and the pinning port. Every uncut edge forces its $q$ endpoint to one. Therefore
\[
B(x,y)=q(1^\ell)^{s-1}q(x,1,\ldots,1)\delta_0(y),
\]
with the singled-out coordinate placed first. In particular $B$ is nonzero of rank one and $B(1,0)\ne0$.

Now replace each formal label in this fixed network by the actual gadget $p(t)$. If $D$ is the degree of $p$, the resulting binary matrix $B(t)$ is a polynomial of degree $rD$ with nonzero leading coefficient $a^rB$. For every positive integer $t$, its normalized tensor belongs to $\mathbb C \NEq\cup\mathbb C \SNEq$. Its diagonal entries consequently vanish identically, and its two off-diagonal polynomials $u(t),v(t)$ satisfy
\[
(u(t)-v(t))(u(t)+v(t))=0
\]
identically. Since $\mathbb C[t]$ has no zero divisors, either $u=v$ identically or $u=-v$ identically. The nonzero leading coefficient must therefore be proportional to $\NEq$ or $\SNEq$, which contradicts its rank one. Equivalently, the displayed rank-one coefficient guarantees a positive integer $t$ for which the actual binary gadget violates the two-class hypothesis. No Turing-equivalent enlargement or formal pinning was used as that contradiction. The other direction follows by interchanging the two bit values throughout the argument.
\end{proof}

\begin{lemma}\label{lem:iso-one-sided-real}\label{thm:iso}
Suppose the transformed binary classes are $[\NEq],[\SNEq]$ with native $\NEq$ edges, and every transformed quaternary gadget is zero, a member of $\mathcal M_4$, or a pure extreme power. If all transformed input signatures have one common one-sided half-weight support, then the corresponding Holant problem is in FP or is \#P-hard.
\end{lemma}
\begin{proof}
For a local assignment to an arity-$n$ vertex, call $|x|-n/2$ its excess. Each internal $\NEq$ edge contributes one one to the sum of local weights. Thus in a gadget with $b$ external ports the sum of local excesses is $|x_{\rm ext}|-b/2$. If the external assignment is balanced, all local excesses of the common sign must be zero. In particular, every closed instance has exactly the same value after every input $\hat f$ is replaced by its middle restriction $\hat f_{=}$.

This also proves the structure of the restricted signature set. For any gadget layout, let $q$ be its tensor over the original inputs and $q^0$ its tensor after middle restriction. The new inputs are balanced, so $q^0$ is balanced too; on balanced boundary assignments the preceding argument gives $q^0=q$. Hence $q^0=q_{=}$. The restricted binary classes are still $[\NEq],[\SNEq]$, and its nonzero quaternary gadgets belong to $\mathcal M_4$: pure extreme powers have zero middle restriction. These assertions follow from the boundary identity, not from complexity equivalence alone.

Return to equality edges and put $f^0=K^{\otimes n}\hat f_{=}$. The restricted signature set now satisfies the $\{\Eq,\SNEq\}$ hypotheses of Lemma~\ref{lem:iso-sharing}. Let
\[
W_n=K^{\otimes n}\{h:\operatorname{supp}(h)\subseteq
\mathrm{HW}^{=}_n\}.
\]
This space is invariant under entrywise conjugation: $\bar K=K\NEq $, and complementing all bits preserves the middle layer. Thus the real and imaginary parts of every vector in $W_n$ also belong to it.

Let $A$ be the real matrix of complete $\Eq,\SNEq$ contractions. The identities $K^TIK=\NEq$ and $K^T\SNEq K=-i\SNEq$ identify these tests, up to nonzero scalars, with the complete middle-layer tests above. Hence $A$ is injective on $W_n$.

For $f^0\ne0$, phase sharing gives $Af^0=\lambda_f r_f$ with $r_f$ real and $\lambda_f\ne0$. Write
\[
f^0/\lambda_f=f_R+ig,\qquad f_R,g\ \hbox{real}.
\]
Both terms lie in $W_n$, and $Ag=0$. Injectivity on $W_n$ gives $g=0$. Normalizing instead by a nonzero entry of $f^0$ makes the scalar and all resulting real entries algebraic. Zero restrictions cause no difficulty: an instance containing one has value zero. We have obtained a finite real algebraic signature set with the same instance values up to known vertex scalars. Apply Theorem~\ref{thm:real-holant} to obtain FP versus \#P-hardness.
\end{proof}

If Lemma~\ref{lem:iso-diagonal-factor} leaves an actual pure extreme quaternary gadget, Lemma~\ref{lem:iso-tail-obstruction} forces a common support direction and Lemma~\ref{lem:iso-one-sided-real} classifies the corresponding Holant problem in FP or as \#P-hard. Henceforth we may work in the residual branch where every quaternary gadget lies in $\mathcal M_4$.

We now use partial contractions to determine a minimum-arity gadget outside $\mathcal M$.
\begin{lemma}[The single-pair kernel]\label{lem:iso-single-kernel}
For a tensor $v$ of arity $m\ge3$, if every single-pair $\NEq$ contraction is zero, then $v=a\delta_{0^m}+b\delta_{1^m}$.
\end{lemma}
\begin{proof}
Take any nonconstant input and choose three positions containing two equal bits and one opposite bit. Let $u_1,u_2,u_3$ be its values when the exceptional bit occupies each of these positions, keeping all other positions fixed. The three zero contractions give $u_1+u_2=u_1+u_3=u_2+u_3=0$. Hence all three values are zero. Every nonconstant input is covered.
\end{proof}

Suppose some actual gadget is outside $\mathcal M$, and choose one, $h$, of minimum arity $n$. Binary and quaternary closure imply $n\ge6$. Every single-pair $\NEq,\SNEq$ contraction of $h$ is in $\mathcal M_{n-2}$, hence is EO. Restriction to the middle layer commutes with these contractions. Therefore $h-h_{=}$ has every single-pair contraction zero, and the preceding lemma gives
\begin{equation}\label{eq:iso-minimum-form}
h=h_{=}+a\delta_{0^n}+b\delta_{1^n}.
\end{equation}
This assertion is for a minimum-arity gadget, not for every input signature. The remainder of the proof determines its middle part.

The next argument specializes the factor-integration method of~\cite[Lemmas 4.1--4.5]{cai2020eoars} to the two factors $\NEq,\SNEq$. Its hypotheses concern the contractions of a tensor, so it also applies to the middle restriction $h_{=}$ before that restriction is realized by a gadget.

\begin{lemma}[Zero merge]\label{lem:iso-zero-merge}
Let $p$ be EO of even arity $n\ge6$, with every single-pair $\NEq,\SNEq$ contraction in $\mathcal M_{n-2}$. If one $\NEq$ contraction is zero, then $p\in\mathcal M_n$.
\end{lemma}
\begin{proof}
Suppose $\partial^{\NEq}_{uv}p=0$. For each disjoint pair $ij$, the product $\partial^{\NEq}_{ij}p$ also has zero $\NEq$ contraction at $uv$. By the product rule above, it is zero or has $(\SNEq)_{uv}$ as a factor. All $\NEq$ contractions of each of the three tensors $p^{00}_{uv},p^{11}_{uv},p^{01}_{uv}+p^{10}_{uv}$ therefore vanish. Their respective weights on $n-2$ positions are $n/2,n/2-2,n/2-1$, none of which is an extreme weight. Lemma~\ref{lem:iso-single-kernel} makes all three tensors zero. Thus $p=(\SNEq)_{uv}\otimes r$. Its $\SNEq$ contraction at $uv$ is $2r$, which belongs to $\mathcal M_{n-2}$, proving the claim.
\end{proof}

\begin{lemma}[High-arity integration]\label{lem:iso-high-integration}
Let $p$ be EO of even arity $n\ge10$. If every single-pair $\NEq,\SNEq$ contraction belongs to $\mathcal M_{n-2}$, then $p\in\mathcal M_n$.
\end{lemma}
\begin{proof}
Assume the contrary. Write $\partial_{ij}=\partial^{\NEq}_{ij}$. By Lemma~\ref{lem:iso-zero-merge} all $\partial_{ij}p$ are nonzero. The hypotheses survive arbitrary local $\SEq$ transformations, because these interchange the two contraction forms up to sign when exactly one contracted port is transformed. Such transformations preserve membership in $\mathcal M$. We can consequently normalize
\[
\partial_{12}p=(\NEq)_{34}(\NEq)_{56}(\NEq)_{78}(\NEq)_{9,10}\cdots (\NEq)_{n-1,n}.
\]
All $\NEq$ merges of the normalized $p$ are again nonzero.

We first establish a propagation rule. Suppose $B_{uv}$ divides both $\partial_{st}p$ and $\partial_{s't'}p$, for distinct pairs $st,s't'$ disjoint from $uv$. Let $ij$ be disjoint from $\{u,v,s,t,s',t'\}$ and suppose both double merges with $ij$ are nonzero. Then $B_{uv}$ divides $R=\partial_{ij}p$. Indeed, if $u,v$ belonged to different binary factors of $R$, a single merge could put them in the same factor only by joining their two partners. That uniquely determines the deleted pair, which cannot be both $st$ and $s't'$. Thus $u,v$ are paired in $R$. This factor survives either double merge; uniqueness of the binary factorization makes it proportional to $B_{uv}$.

Next suppose $B_{uv}$ divides a triangle $\partial_{rs}p,\partial_{rt}p,\partial_{st}p$, with $u,v,r,s,t$ distinct. For $ij$ disjoint from all five indices, at most one of the three triangle merges of $\partial_{ij}p$ is zero: a zero would require a $\SNEq$ factor on that pair, and two overlapping pairs cannot both be factors. The propagation rule therefore gives $B_{uv}\mid\partial_{ij}p$. If $ij$ is disjoint from $uv$ but intersects the triangle, choose three new indices outside $uv\cup ij\cup\{r,s,t\}$. This union has at most six elements, so $n\ge10$ suffices. The disjoint case gives a new triangle with the same factor; apply that case once more to $ij$. Hence $B_{uv}$ divides every $\partial_{ij}p$ disjoint from $uv$.

This common divisibility integrates to $B_{uv}\mid p$. Write the entries of $B$ at $01,10$ as $\beta,\gamma\ne0$. Every $\NEq$ contraction of $p^{00}_{uv},p^{11}_{uv},\gamma p^{01}_{uv}-\beta p^{10}_{uv}$ is zero. Their supports avoid the extreme weights on $n-2$ positions. Lemma~\ref{lem:iso-single-kernel} makes them zero, giving the asserted factorization.

It remains to find the triangle. The nonzero product $R=\partial_{34}p$ satisfies
\[
\partial_{12}R=2(\NEq)_{56}(\NEq)_{78}(\NEq)_{9,10}\cdots (\NEq)_{n-1,n}.
\]
If $1,2$ are paired in $R$, all its other pairs are the displayed ones. Otherwise their two partners form exactly one displayed pair, which we may name $5,6$. In either case $(\NEq)_{78}$ and $(\NEq)_{9,10}$ are factors of $R$, and $6$ is paired with neither $7$ nor $8$. Thus the merges $67,68,78$ of both $\partial_{12}p$ and $\partial_{34}p$ are nonzero, with $(\NEq)_{9,10}$ preserved. The propagation rule gives
\[
(\NEq)_{9,10}\mid\partial_{67}p,\quad
(\NEq)_{9,10}\mid\partial_{68}p,\quad
(\NEq)_{9,10}\mid\partial_{78}p.
\]
The triangle and integration arguments give $(\NEq)_{9,10}\mid p$. Contracting this factor by $\NEq$ isolates its complement, which is in $\mathcal M_{n-2}$ by hypothesis. This contradicts $p\notin\mathcal M_n$.
\end{proof}

Index eight positions by $t\in\mathbb F_2^3$, in binary order, and define
\[
C=\{(a_0+a_1t_1+a_2t_2+a_3t_3)_{t\in\mathbb F_2^3}:
          a\in\mathbb F_2^4\},\qquad H=\mathbf1_C.
\]
This is $\mathrm{RM}(1,3)$. Its occurring weights are $0,4,8$, and $H_{=}$ has fourteen support entries.

\begin{lemma}[Exact six- and eight-port classification]
\label{lem:iso-finite}
Let $p$ be EO of arity $n\in\{6,8\}$, with every single-pair $\NEq,\SNEq$ contraction in $\mathcal M_{n-2}$. For $n=6$, $p\in\mathcal M_6$. For $n=8$, either $p\in\mathcal M_8$, or, up to a nonzero scalar, a port permutation and local $\SEq$ transformations, $p=H_{=}$.
\end{lemma}
\begin{proof}
The zero tensor is immediate, and Lemma~\ref{lem:iso-zero-merge} handles a zero unsigned contraction. Otherwise normalize
\[
\partial^{\NEq}_{12}p=P:=(\NEq)_{34}(\NEq)_{56}\cdots(\NEq)_{n-1,n}.
\]
Local signed equalities on ports other than $1,2$ give this normalization and preserve the hypotheses.

For a nonzero product $R=\lambda\prod_{uv}C_{uv}$ of signed or unsigned disequalities, define its product preimages by
\[
\mathscr L_{\neq}(R)=
\left\{\frac12(\NEq)_{12}R\right\}
\ \cup\ 
\left\{\lambda B_{1u}(\NEq B^{-T}C)_{2v}
             \prod_{ab\ne uv}C_{ab}\right\}.
\]
Choose a factor $C_{uv}$, either order of its two ports, and $B\in\{\NEq,\SNEq\}$. The contraction identity $B^T\NEq(\NEq B^{-T}C)=C$ proves that every listed product contracts to $R$. Conversely, either $1,2$ are paired together, or their partners form one factor of $R$; these alternatives prove completeness of the list. Its size is $1+2\,\operatorname{arity}(R)$.

Let $A_{ij}^{\pm}$ act on the middle-layer coordinates of $p$: its row for $y$ has coefficients $1,\pm1$ at the two words obtained by inserting $01,10$ in positions $i,j$. Thus $A_{ij}^+p=\partial^{\NEq}_{ij}p$ and $A_{ij}^-p=\partial^{\SNEq}_{ij}p$. Commutativity gives the finite affine conditions
\[
A_{12}^+p=P,\qquad
A_{ij}^+p\in\mathscr L_{\neq}(A_{ij}^+P)
\quad(3\le i<j\le n).
\]
Indices on $P$ are inherited after deleting $1,2$. The images $A_{ij}^+P$ are nonzero, so each list has five members at arity six and nine at arity eight, with fixed scalars. Eliminate coordinates in lexicographic order, starting from $A_{12}^+p=P$. At each pair, append each listed right-hand side, reduce by the preceding rows, and retain it exactly when every zero row has zero right-hand side. The ranks and numbers of distinct consistent right-hand sides are
\[
\begin{array}{c|rrrrrr}
n=6:\ ij&34&35&36&45&46&56\\ \hline
\text{rank}&10&14&17&20&20&20\\
\text{number}&5&25&25&125&57&17
\end{array}
\qquad
\begin{array}{c|rrrrrr}
n=8:\ ij&34&35&36&37&38&45\\ \hline
\text{rank}&34&48&58&64&67&70\\
\text{number}&9&81&181&25&17&17
\end{array}
\]
At these final ranks, every retained system is a single point. The remaining disjoint pairs at arity eight keep the same seventeen points. Impose the original conditions $A_{ij}^{\pm}p\in\mathcal M_{n-2}$ for all pairs $ij$. Direct substitution gives
\[
\begin{array}{c|c|l}
n&\text{remaining points}&\text{solution set}\\ \hline
6&9&\mathscr L_{\neq}(P)\\
8&17&\mathscr L_{\neq}(P)\ \cup\ \mathscr E
\end{array}
\]
where
\[
\mathscr E=\{H_{=},\ \tau H_{=},\
  -(\SEq)_1(\SEq)_2H_{=},\
  -(\SEq)_1(\SEq)_2\tau H_{=}\},
\]
and $\tau$ exchanges the last two ports. Every row and listed right-hand side is rational. The elimination therefore classifies complex solutions as well as rational ones.

The product part has $2n-3$ points, namely nine and thirteen. Each of the four additional eight-port tensors has the stated normalization; its signed and unsigned contractions are matching products, by the affine-linear description of $C$. Conversely, every tensor under consideration satisfies the affine conditions just classified, so it belongs to one of the displayed sets. This proves the two assertions, including the signed and permuted form of every eight-port exception.
\end{proof}

Combining these lemmas with~\eqref{eq:iso-minimum-form}, the minimum gadget has one of the following two forms after actual local $\SEq$ transformations, a port permutation, and scalar normalization:
\begin{align}
h&=cP+a\delta_{0^n}+b\delta_{1^n},\qquad
             P=(\NEq)_{1,d+1}\cdots (\NEq)_{d,2d},\ n=2d;\label{eq:iso-product-tail}\\
h&=H_{=}+a\delta_{0^8}+b\delta_{1^8}.\label{eq:iso-rm-tail}
\end{align}
The first form includes $h_{=}=0$. We resolve these two forms separately.

\begin{lemma}\label{lem:iso-unipotent}
An actual signature of form~\eqref{eq:iso-product-tail} outside $\mathcal M$, in the residual branch, gives the dichotomy for $\Holant(\F)$ asserted in Theorem~\ref{thm:iso-complete}.
\end{lemma}
\begin{proof}
First $ab=0$. In two copies leave one complete factor pair of $P$ open in each copy and connect every other port to its corresponding port in the other copy, using $\NEq$ except for one $\SNEq$ link. Any middle--tail contribution is zero because an internal $P$ pair requires opposite bits while a tail fixes both. The middle--middle contribution is also zero: the component with one $\SNEq$ link is a closed binary cycle with zero trace. The two opposite-tail contributions give, up to a nonzero sign,
\[
ab(\delta_{0011}-\delta_{1100}).
\]
If nonzero, this four-port tensor has support size two, whereas every nonzero member of $\mathcal M_4$ has support size four. Thus $ab=0$.

If $c=0$, the nonzero gadget is a pure extreme power. Lemma~\ref{lem:iso-tail-obstruction} forces a common support direction, and Lemma~\ref{lem:iso-one-sided-real} gives FP versus \#P-hardness. If $c\ne0$, normalize $c=1$. Since $h\notin\mathcal M$, precisely one tail coefficient is nonzero; by symmetry suppose it is $a$. Use the displayed left--right partition of $P$ and put $E=(\NEq)^{\otimes d}$. The transfer matrix is
\[
M_hE=\mathrm{Id}+aT,\qquad
T=|0^d\rangle\langle1^d|,\qquad T^2=0.
\]
Joining $k$ copies in series through $d$ $\NEq$ edges at each step therefore realizes
\[
h_k=P+ka\delta_{0^n}\qquad(k\ge1).
\]
The leading coefficient of this actual polynomial family is $a\delta_{0^n}$. Lemma~\ref{lem:iso-tail-obstruction} therefore forces every input to have support below or at half weight. In particular the tail disappears from every closed evaluation by weight conservation. Lemma~\ref{lem:iso-one-sided-real} makes the whole surviving signature set real up to scalars and gives FP versus \#P-hardness. The $b$-tail case gives the reversed common direction.
\end{proof}

The bounded-occurrence method is inspired by Section 8.2 of the full version of the real Holant proof~\cite{shao2020dichotomy}. We give the resource identities and the additional phase argument needed for the present complex-valued signature set explicitly.
\begin{lemma}\label{lem:iso-rm-terminal}
An actual signature of form~\eqref{eq:iso-rm-tail}, in the residual branch, implies that $\Holant(\F)$ is in FP or is \#P-hard.
\end{lemma}
\begin{proof}
Use two copies of $h$. Leave the first two ports of each copy open, and join the remaining six pairs of corresponding ports by $\NEq$, except that the first of these six links is $\SNEq$. The resulting tensor is exactly
\begin{equation}\label{eq:iso-rm-mating}
(ab-1)(\delta_{0011}-\delta_{1100}).
\end{equation}
Here is a direct coefficient check. Puncturing the first two coordinates is injective on $C$, so matched words must be complements on all eight coordinates. For the first two bits $00$, the three middle words have signs summing to $-1$ at the marked internal coordinate; the extreme word contributes $ab$. For first bits $01$ or $10$, the four middle words have two signs of each kind. Complementation reverses the marked sign, giving the negative coefficient for $11$. This proves the displayed identity. Residual four-port closure forces $ab=1$.

Choose an algebraic $z\ne0$ with $z^8=a^{-1}$ and apply $D_z=\diag(z,z^{-1})$ to the entire right-hand signature set. Since $D_z^T\NEq D_z=\NEq$, the edge kernel is unchanged, and the binary classes remain $[\NEq],[\SNEq]$. The eight-port gadget becomes exactly $H$: its middle layer is unchanged, while its endpoint values are $az^8=bz^{-8}=1$. This is a common holographic change of basis, corresponding to the orthogonal matrix $KD_zK^{-1}$ in the original presentation.

Return to ordinary equality edges and put $R=K^{\otimes8}H$. Since $\bar K=K\NEq$ and $H$ is invariant under complementation, $R$ is real. The preceding common basis change is orthogonal in these coordinates. It preserves $I$ and the class of $\SNEq$, so the binary and quaternary hypotheses of Lemma~\ref{lem:iso-sharing} still hold.

We use one additional real contraction form, $\SEq$, and eliminate it by the bounded-occurrence method of~\cite[Section 8.2]{shao2020dichotomy}. The required identities are
\begin{equation}\label{eq:iso-rm-replicator}
\partial^{\SEq}_{12}R=(\SEq)_{34}(\SEq)_{56}(\SEq)_{78},\qquad
\big\langle(\Eq)_{12}(\Eq)_{34}(\Eq)_{56}(\Eq)_{78},R\big\rangle=8.
\end{equation}
For the first identity, $K^T\SEq K=I$ reduces the contraction to imposing equality on coordinates $000,001$ of an affine word in $C$. This sets one linear coefficient to zero. The remaining six coordinates are three equal pairs with three independent values, giving $I\otimes I\otimes I$ in the $K$ basis. Since $KIK^T=\SEq$, the identity follows. For the second, $K^TIK=\NEq$ fixes that same linear coefficient to one and leaves the other three affine parameters free.

Replacing three auxiliary $\SEq$ vertices by one $R$ and one $\SEq$ decreases their number by two. Repetition leaves at most two. With one auxiliary removed, the remaining binary belongs to $\{\Eq,\SNEq\}$ up to scalar and is orthogonal to $\SEq$, so the value is zero. With two removed, the remaining quaternary $q$ satisfies
\begin{equation}\label{eq:iso-real-two-J}
\langle(\SEq)_{12}(\SEq)_{34},q\rangle
=\tfrac12\langle(\Eq)_{13}(\Eq)_{24},q\rangle
-\tfrac12\langle(\SNEq)_{13}(\SNEq)_{24},q\rangle.
\end{equation}
To verify this, write $q=K^{\otimes4}\hat q$. The residual quaternary hypothesis gives $\hat q\in\mathcal M_4$, so it is balanced. In these coordinates both sides test $\hat q(0011)+\hat q(1100)$: the crossing $\NEq$ and $\SNEq$ tests cancel the other two entries. The minus sign comes from $K^T\SNEq K=-i\SNEq$ in each of the two signed contractions. Thus all auxiliary eliminations have real coefficients. Binary-only components reduce to real matrix traces and can be evaluated separately.

Fix a nonzero original-basis input $f$ of arity $2d$. Consider its complete contractions using the four real forms $\Eq,\SEq,\NEq,\SNEq$ on one fixed pairing. The extra forms here are analytic tests, not signatures assumed available. Since $\NEq=\SEq \SNEq $, any such test is a closed network with one $f$, at most $d$ auxiliary $\SEq$ vertices, and ordinary $\Eq,\SNEq$ vertices. Replace three $\SEq$ vertices by one $R$ and one $\SEq$ repeatedly, and then use the zero rule or~\eqref{eq:iso-real-two-J} for the at most two remaining auxiliaries. Each test value is thus a real linear combination of complete $\Eq,\SNEq$ networks with one $f$ and $k\le d$ copies of $R$.

Pad every such network by $d-k$ disconnected copies of the fixed $R$ closure of value $8$, compensating by the real factor $8^{-(d-k)}$. All query values now arise from the same actual tensor
\[
v_f=f\otimes R^{\otimes d}
\]
by complete $\Eq,\SNEq$ contractions, up to real factors. By Lemma~\ref{lem:iso-sharing}, their nonzero values lie in one complex line $\lambda_f\mathbb R$. At least one is nonzero: otherwise every four-form test of $f$ would be zero, contradicting the fact that $\Eq,\SEq,\NEq,\SNEq$ span all real binary tensors. Hence all four-form tests of $f$ lie in $\lambda_f\mathbb R$ with $\lambda_f\ne0$.

Their real coefficient matrix $A$ has full column rank on the entire arity-$2d$ tensor space, so
\[
A\operatorname{Im}(f/\lambda_f)=0
\quad\Longrightarrow\quad
\operatorname{Im}(f/\lambda_f)=0.
\]
Apply this to every input $f$ in the same orthogonal basis. Lemma~\ref{lem:real} and the real Holant dichotomy classify the whole finite signature set.
\end{proof}

\begin{proof}[Proof of Theorem~\ref{thm:iso-complete}]
Apply Lemma~\ref{lem:iso-diagonal-factor}. It either classifies the Holant problem by the real Pauli extension, or leaves only matching products and pure extreme quaternary gadgets. If a pure extreme occurs, Lemmas~\ref{lem:iso-tail-obstruction} and~\ref{lem:iso-one-sided-real} classify $\Holant(\F)$ in FP or as \#P-hard. Otherwise we are in the residual branch. If every input belongs to $\mathcal M$, split its fixed binary factors and evaluate the resulting paths and cycles in polynomial time. Otherwise an actual gadget outside $\mathcal M$ exists; choose one of minimum arity. All its lower-arity contractions belong to $\mathcal M$, giving~\eqref{eq:iso-minimum-form}. For $n\ge10$ use Lemma~\ref{lem:iso-high-integration} on $h_{=}$. For $n=6,8$ use Lemma~\ref{lem:iso-finite}. Thus the minimum gadget has form~\eqref{eq:iso-product-tail} or~\eqref{eq:iso-rm-tail}. Lemmas~\ref{lem:iso-unipotent} and~\ref{lem:iso-rm-terminal} classify $\Holant(\F)$ in the respective cases. The matching-product case is in FP; each remaining branch reduces the entire signature set to the real Holant dichotomy.
\end{proof}

\section{The indecomposable quaternary signatures}
\label{sec:5}
Assume that the gadget closure contains a tensor-indecomposable quaternary signature. We classify the normal and isotropic cases by gadget constructions and interpolation. Every reduction applies to $\Holant(\F)$ with all signatures in $\F$ retained.

\subsection{Structural Preliminary}
Adjoin one fixed indecomposable quaternary gadget and denote it by $f$. This fixed-gadget substitution preserves the original gadget closure as well as the complexity of the problem.

In the subsequent discussion, we will frequently represent a quaternary signature as a $4\times 4$ signature matrix. For a quaternary signature $f$, following the notation introduced above, we denote its signature matrix by $M(f)_{x_1x_2,x_3x_4}$, or simply by $M(f)$ (or $M_f$) when no confusion arises. The matrix $M(f)$ takes the following form:
\[
M_{x_1x_2,x_3x_4}(f) =\left( 
\begin{matrix}
f(0,0,0,0) & f(0,0,0,1) & f(0,0,1,0) & f(0,0,1,1) \\
f(0,1,0,0) & f(0,1,0,1) & f(0,1,1,0) & f(0,1,1,1) \\
f(1,0,0,0) & f(1,0,0,1) & f(1,0,1,0) & f(1,0,1,1) \\
f(1,1,0,0) & f(1,1,0,1) & f(1,1,1,0) & f(1,1,1,1)
\end{matrix}\right).
\]
For notational convenience, we also allow variables to be written as subscripts, i.e., \(F_{ijkl}=f(i,j,k,l)\) with \(i,j,k,l\in\{0,1\}\) and when there is no risk of confusion, we may omit $f$ and simply write $M_{x_i x_j, x_k x_\ell}$ when describing the matrix.

We use two gadget constructions: attaching a binary signature to one port of a quaternary signature, and connecting two quaternary signatures along two ports. 

In the first case, given a binary signature $B(x,x')$ and a quaternary signature $F(x_1,x_2,x_3,x_4)$, we connect one variable of $F$ with one variable of $B$ to obtain a new quaternary signature $F'$ from $F$. This construction can be described by the following formula:
\[
F'(x_1,x_2,x_3,x_4) = \sum_{x_4'\in\{0,1\}} F(x_1,x_2,x_3,x_4') B(x_4',x_4) = \pij{56}{=_2}(F(x_1,x_2,x_3,x_5)\otimes B(x_6,x_4)).
\]
In the second case, order the shared ports last in $G_1(x_1,x_2,y_1,y_2)$ and first in $G_2(y_3,y_4,x_3,x_4)$. Connecting $y_1$ to $y_3$ and $y_2$ to $y_4$ gives
\[
\begin{aligned}
G'(x_1, x_2, x_3, x_4) &= \sum_{a, b \in \{0,1\}} G_1(x_1, x_2, a, b) \; G_2(a, b, x_3, x_4)\\ 
& = \pij{56}{=_2}\pij{78}{=_2}(G_1(x_1,x_2,x_5,x_7)\otimes G_2(x_6,x_8,x_3,x_4)). 
\end{aligned}
\]

\begin{figure}[H]
    \centering
    \begin{minipage}{0.45\textwidth}
        \centering
        \begin{tikzpicture}[scale=0.9]
            \node[draw, rectangle, rounded corners, minimum width=4cm, minimum height=3cm] (box) {};
            \coordinate (p1) at ($(box.center) + (-0.9,0)$);
            \coordinate (p2) at ($(box.center) + (0.8,0)$);
            \filldraw (p1) circle (2pt) node[below left] {\large $F$};
            \filldraw (p2) circle (2pt) node[below right] {\large $B$};
            \node[draw, dashed, fit=(p1), inner sep=18pt] {};
            \node[draw, dashed, fit=(p2), inner sep=18pt] {};
            \draw ($(box.north west)!0.3!(box.north east)$) -- ++(0,0.8);
            \draw ($(box.south west)!0.3!(box.south east)$) -- ++(0,-0.8);
            \draw (box.west) -- ++(-0.8,0);
            \draw (box.east) -- ++(0.8,0);
            \draw (p1) -- ($(box.north west)!0.3!(box.north east)$);
            \draw (p1) -- ($(box.south west)!0.3!(box.south east)$);
            \draw (p1) -- (box.west);
            \draw (p2) -- (box.east);
            \draw (p1) -- (p2);
            \node[below=0.2cm of box.south] {$F'$};
        \end{tikzpicture}
            
    \end{minipage}
    \hfill
    \begin{minipage}{0.45\textwidth}
        \centering
        \begin{tikzpicture}[scale=0.9]
            \node[draw, rectangle, rounded corners, minimum width=4cm, minimum height=3cm] (box) {};
            
            \coordinate (p1) at ($(box.center) + (-0.85,0)$);
            \coordinate (p2) at ($(box.center) + (0.85,0)$);
            \filldraw (p1) circle (2pt) node[left] {\large $G_1$};
            \filldraw (p2) circle (2pt) node[right] {\large $G_2$};
            \node[draw, dashed, fit=(p1), inner sep=18pt] {};
            \node[draw, dashed, fit=(p2), inner sep=18pt] {};
            \draw ($(box.north west)!0.3!(box.north east)$) -- ++(0,0.8);
            \draw ($(box.north west)!0.7!(box.north east)$) -- ++(0,0.8);
            \draw ($(box.south west)!0.3!(box.south east)$) -- ++(0,-0.8);
            \draw ($(box.south west)!0.7!(box.south east)$) -- ++(0,-0.8);
            \draw (p1) -- ($(box.north west)!0.3!(box.north east)$);
            \draw (p1) -- ($(box.south west)!0.3!(box.south east)$);
            \draw (p2) -- ($(box.north west)!0.7!(box.north east)$);
            \draw (p2) -- ($(box.south west)!0.7!(box.south east)$);
            \draw (p1) to[bend left] (p2);
            \draw (p1) to[bend right] (p2);
        \node[below=0.2cm of box.south] {$G'$};
        \end{tikzpicture}
        \centering
    \end{minipage}
    \caption{Attaching a binary signature $B$ to one port of $F$ gives $F'$; coupling $G_1$ and $G_2$ along two ports gives $G'$.}
    \label{fig:non_gad}
\end{figure}
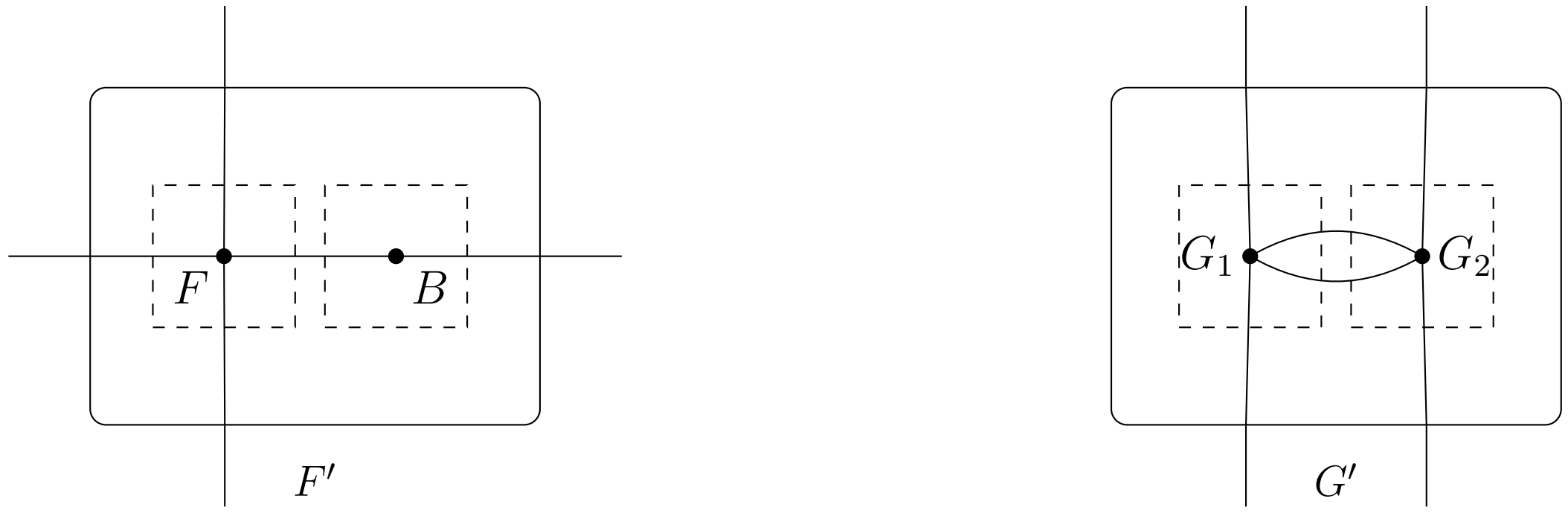

These formulas express the same constructions as explicit sums and as contractions. With the indicated port order, the second construction is matrix multiplication. Both are illustrated in Figure~\ref{fig:non_gad}.

For instance, we can construct a quaternary signature \(G\) by coupling two copies of \(F\) (with the variables of one copy suitably reordered) along two shared variables, so that \(M(G) = M(F)^2\). Concretely, we obtain \(G(x_1,x_2,x_5,x_6)\) by connecting the variables \(x_3,x_4\) of \(F(x_1,x_2,x_3,x_4)\) with the same variables of \(F(x_3,x_4,x_5,x_6)\). The identity
\[
M_{x_1x_2,x_5x_6}(G) = M_{x_1x_2,x_3x_4}(F)\, M_{x_3x_4,x_5x_6}(F)
\]
follows immediately from the formula above and the definition of matrix multiplication.

Changing the row and column ports gives the following two matrices for the same tensor: { \setlength{\arraycolsep}{3.8pt}
\[
M_{x_1x_3,\,x_2x_4} =
\begin{pmatrix}
F_{0000} & F_{0001} & F_{0100} & F_{0101} \\
F_{0010} & F_{0011} & F_{0110} & F_{0111} \\
F_{1000} & F_{1001} & F_{1100} & F_{1101} \\
F_{1010} & F_{1011} & F_{1110} & F_{1111}
\end{pmatrix},
M_{x_1x_4,\,x_2x_3} =
\begin{pmatrix}
F_{0000} & F_{0010} & F_{0100} & F_{0110} \\
F_{0001} & F_{0011} & F_{0101} & F_{0111} \\
F_{1000} & F_{1010} & F_{1100} & F_{1110} \\
F_{1001} & F_{1011} & F_{1101} & F_{1111}
\end{pmatrix}.
\]}
The six length-four strings of weight two form three complementary pairs. A permutation of the ports can permute these three pairs arbitrarily. This relabels the tensor's arguments and is an allowed gadget operation; it need not fix the ordered tensor entrywise.

The displayed multiplication formulas use ordinary equality edges. With edge form $\NEq$, coupling two quaternary signatures instead has matrix $M_1(\NEq\otimes \NEq)M_2$. This distinction is retained throughout the isotropic proof.

The same elementary algebraic observation will resolve the residual square equations in both cases.
\begin{lemma}\label{lem:paired-squares}
For arbitrary $S,A,C,E\in\mathbb C$, the identities
\[
(S-E)^2=(A-C)^2,\qquad
(S-A)^2=(C-E)^2,\qquad
(S-C)^2=(E-A)^2
\]
hold if and only if
\[
(S=E,\ A=C),\quad (S=A,\ C=E),\quad\text{or}\quad(S=C,\ A=E).
\]
The two pairs may have the same value, and zero values are allowed.
\end{lemma}
\begin{proof}
Write $L_1$ for $S+C=A+E$, $L_2$ for $S+A=C+E$, and $L_3$ for $S+E=A+C$. Factoring each difference of squares gives, in order, $(L_1\text{ or }L_2)$, $(L_1\text{ or }L_3)$, and $(L_2\text{ or }L_3)$. Thus at least two of the $L_i$ hold. The pairs $(L_1,L_2)$, $(L_1,L_3)$, and $(L_2,L_3)$ give the three displayed alternatives, respectively. Conversely each alternative satisfies all three squared identities by substitution.
\end{proof}

The following interpolation applies to either edge convention and will be used in both cases.
\begin{lemma}[Projector interpolation]
\label{lem:trueq-projector}
Let $J\in\{I,\NEq\}$ be the edge matrix and let $T=M(h)(J\otimes J)$ be the transfer matrix of an available quaternary signature. Suppose $T$ is zero on one invariant subspace and invertible on its invariant complement $W$, with at most two nonzero eigenvalues and Jordan blocks of size at most two. Let $P$ be the projector onto $W$ along the zero subspace. Then the signature with matrix $P(J\otimes J)$ can be adjoined by a polynomial-time Turing reduction.
\end{lemma}
\begin{proof}
A chain of $k\geq1$ copies has transfer matrix $T^k$. If $W=\{0\}$, the target signature is zero and any instance containing it has value zero. Assume henceforth that $W\ne\{0\}$. If there are no target occurrences, no replacement is needed. Otherwise let $m\ge1$ be their number. Every positive power of $T$ vanishes on the zero subspace, as does $P$. Thus the target value and all chain query values are obtained from their restrictions to $W$, with exactly the same contraction against the rest of the instance. If the nonzero eigenvalues are $\alpha,\beta$ (repeating one if necessary), then
\[
p(z)=\prod_{j=0}^{m}
  (z-\alpha^j\beta^{m-j})^{m+1}
\]
annihilates the tensor product of these $m$ restricted transfers. Indeed, a tensor product of $m$ Jordan blocks of size at most two has nilpotent index at most $m+1$ after its eigenvalue is subtracted. Write $p(z)=\sum_{k=0}^{D}p_kz^k$; its constant coefficient is nonzero. After contraction with the unchanged rest of the instance, the identity $p((T|_W)^{\otimes m})=0$ gives $p_0 Z_0+\sum_{k=1}^{D}p_kZ_k=0$. Here $Z_0$ is the target value, and $Z_k$ is obtained by replacing every target occurrence with a length-$k$ chain. This recovers $Z_0$ using $D\leq(m+1)^2$ queries of polynomial size. Constructing $p$ uses only its two fixed algebraic eigenvalues and polynomial-size coefficient arithmetic; no exponentially large tensor-product matrix is constructed. Repeated eigenvalues and roots of unity cause no exception.
\end{proof}

\subsection{Normal case}
\label{ssec:trueq-normal}
\label{ssec:normal5}
The available binary classes are $[\Eq],[\SEq]$. By Lemma~\ref{lem:odd}, every signature has even parity. Write the matrix of the chosen indecomposable quaternary gadget as
\[
M(h)=\begin{pmatrix}s&0&0&e\\0&a&c&0\\0&d&b&0\\f&0&0&t\end{pmatrix}.
\]
For each complementary middle pair $(u,v)\in\{(a,b),(c,d),(e,f)\}$, the two corresponding choices of contracted ports, with labels $\Eq,\SEq$, give
\[
(s+u)^2=(v+t)^2,\quad(s-u)^2=(v-t)^2,\qquad
(s+v)^2=(u+t)^2,\quad(s-v)^2=(u-t)^2.
\]
These equations hold because the resulting binary is zero or has two diagonal entries equal up to sign. Adding and subtracting gives
\begin{equation}\label{eq:normal-complementary}
s^2=t^2,\qquad u^2=v^2,\qquad su=tv,\qquad sv=tu.
\end{equation}
Thus $st=0$ means $s=t=0$. If $st\ne0$, every nonzero complementary pair, including $(s,t)$, has the same ratio $s/t\in\{1,-1\}$.

\begin{theorem}
Let $\F$ be a finite set of even-arity signatures. If its nonzero binary gadget classes are $[\Eq],[\SEq]$ and an indecomposable quaternary signature is realizable from $\F$, then $\Holant(\F)$ is in FP or is \#P-hard.
\label{thm:tnm}
\end{theorem}
\begin{proof}
We construct $=_4$, possibly after a common holographic transformation, and then apply Theorem~\ref{thm:csp-even} to the whole transformed signature set. There are two cases according to the extreme entries.

Suppose first that $s=t=0$, and let $r$ be the number of nonzero middle complementary pairs. Since $h\ne0$, we have $r\ge1$. If $r=1$, place the pair at $(e,f)$. Then $M(h)^2=ef\,\diag(1,0,0,1)$ supplies $=_4$. If $r=2$, place the pairs in the middle block $B=\left(\begin{smallmatrix}a&c\\d&b\end{smallmatrix}\right)$. A singular $B$ would express $h$ as a tensor product on $12\mid34$, so $B$ is invertible. Lemma~\ref{lem:trueq-projector}, with $J=I$, supplies $\diag(0,1,1,0)$. Move its two support strings to $0011,1100$ and square to obtain $=_4$.

For $r=3$, two complementary ratios agree. Permute the middle pairs and, if necessary, attach $\SEq$ at the first port so that $b=-a,d=-c$. Write $f=\eta e$, where $\eta\in\{1,-1\}$. Squaring gives
\[
\diag(\eta e^2,a^2-c^2,a^2-c^2,\eta e^2).
\]
If $e^4\ne(a^2-c^2)^2$, Lemma~\ref{lem:intp} supplies $=_4$. Otherwise attach $\SEq$ at the first port of the unsquared signature and square to obtain
\[
M(g)=\begin{pmatrix}
-\eta e^2&0&0&0\\
0&a^2+c^2&2ac&0\\
0&2ac&a^2+c^2&0\\
0&0&0&-\eta e^2
\end{pmatrix}.
\]
This is an indecomposable gadget with both extreme entries nonzero. Indeed, if $a^2+c^2\ne0$, it has six support strings. An even-parity tensor with both extreme entries nonzero has no unary factor, and any binary tensor factors have fixed parity, so their product has at most four support strings. If $a^2+c^2=0$, the only possible binary factorization is across the equal-bit pairs of its four support strings; its coefficient determinant is $e^4-(2ac)^2=-8a^2c^2\ne0$. Thus $g$ is covered by the following case.

Now suppose $st\ne0$. By~\eqref{eq:normal-complementary}, a local $\SEq$ makes all complementary ratios $-1$. Set $S=s^2,A=a^2,C=c^2,E=e^2$. Squaring with the ordered row--column partitions $12\mid34$, $13\mid24$, and $14\mid32$ gives
\[
\begin{split}
D_1&=\diag(S-E,A-C,A-C,S-E),\\
D_2&=\diag(S-A,E-C,E-C,S-A),\\
D_3&=\diag(S-C,A-E,A-E,S-C).
\end{split}
\]
If any of their $2\times2$ coefficient tables has nonzero determinant, Lemma~\ref{lem:intp} supplies $=_4$. Otherwise Lemma~\ref{lem:paired-squares} allows a permutation of the middle pairs with $E=S$ and $C=A$. Return the common complementary ratio to $+1$. The signature now has outer block $sU_\delta$ and middle block $aU_\gamma$, where
\[
U_\xi=\begin{pmatrix}1&\xi\\\xi&1\end{pmatrix},\qquad
e=\delta s,\quad c=\gamma a,\quad\delta,\gamma\in\{1,-1\}.
\]
If $a=0$, the sole block has rank one, so $h$ decomposes. Hence $a\ne0$. Since $U_\xi^2=2U_\xi$, Lemma~\ref{lem:trueq-projector} with $J=I$, followed by multiplication by two, gives blocks $U_\delta,U_\gamma$. Attaching $\SEq$ at ports 1 and 3 when $\delta=-1$ makes them $U_1,U_\eta$, where $\eta=\delta\gamma$.

For $\eta=1$, this is the symmetric even-parity signature $[1,0,1,0,1]$. The orthogonal matrix $H_2=2^{-1/2}\left(\begin{smallmatrix}1&1\\1&-1\end{smallmatrix}\right)$ satisfies
\[
H_2^{\otimes4}[1,0,1,0,1]=2(=_4).
\]
For $\eta=-1$, attach $\SEq$ at ports 1 and 4 to obtain $[1,0,-1,0,1]=2K^{\otimes4}(=_4)$. Transforming the whole problem by $K^{-1}$ supplies $=_4$ with $\NEq$ edges. The original binaries become
\[
K^{-1}\SEq(K^{-1})^T=I,\qquad K^{-1}I(K^{-1})^T=\NEq.
\]
An $I$ vertex between two $\NEq$ edges supplies an equality wire, since $\NEq I\NEq=I$; conversely an ordinary $\NEq$ vertex supplies a disequality edge. We therefore obtain an ordinary Holant problem with $=_4$ also in this case.

Every reduction retains all input signatures. In each case the resulting problem is $\Holant(\G,=_4)\equiv_T\CSP_2(\G)$ for the entire original or transformed set $\G$, including the specified binaries. Theorem~\ref{thm:csp-even} gives FP when $\G$ lies in one common class among $\Prod,\mathcal A,\mathcal A_2^1,\mathcal L$, and \#P-hardness otherwise.
\end{proof}

\subsection{Isotropic case}
\label{ssec:trueq-isotropic}
We use the coordinates of Section~\ref{ssec:isotropic}: the edge form is $\NEq$, and the available right-hand binaries are $\NEq,\SNEq$. Attaching $\SNEq$ implements a local $\SEq$ up to sign. As in Section~\ref{ssec:trueq-normal}, two-copy gadgets and projector interpolation reduce the classification to equality signatures. Here the intermediate equality is balanced; we first prove that adjoining it preserves the binary classes, and then classify the entire signature set containing it.

By Lemma~\ref{lem:iso-parity4}, a realizable quaternary signature has matrix
\[
M(h)=\begin{pmatrix}
s&0&0&e\\0&a&c&0\\0&d&b&0\\r&0&0&t
\end{pmatrix}.
\]
The six choices of a pair, contracted with $\NEq$ and $\SNEq$, give
\begin{equation}\label{eq:epsilon}
a^2=b^2,\quad c^2=d^2,\quad e^2=r^2,\qquad
ac=bd,\ ad=bc,\ ae=br,\ ar=be,\ ce=dr,\ cr=de.
\end{equation}
For example, two of these pairs give $(e+c)^2=(d+r)^2$, $(e-c)^2=(d-r)^2$, $(e+d)^2=(c+r)^2$, and $(e-d)^2=(c-r)^2$. Adding and subtracting these equations and their variable permutations gives~\eqref{eq:epsilon}. Thus all nonzero middle complementary pairs have one common sign: $b=\epsilon a$, $d=\epsilon c$, $r=\epsilon e$, where $\epsilon\in\{1,-1\}$. A local $\SEq$ makes $\epsilon=1$. This normalization concerns the chosen signature only. Neither a common sign for the signature set nor a relation between $s$ and $t$ is asserted.

We first establish the reductions needed for $\Holant(\NEq\mid\G)$, where $\G$ denotes the current set of right-hand signatures.
\begin{corollary}\label{cor:projector-binary-closure}
For $J=\NEq$, if the original nonzero binary classes are $[\NEq],[\SNEq]$, adjoining the projector signature of Lemma~\ref{lem:trueq-projector} preserves these binary classes.
\end{corollary}
\begin{proof}
If the projector is zero, any gadget containing it is zero, so adjoining it preserves the nonzero binary classes. Otherwise, fix any binary gadget using $m$ occurrences of the new signature. The case $m=0$ is the original binary hypothesis, so assume $m\ge1$. Replace all of them by length-$k$ chains and call the resulting matrix $B_k$, with $B_0$ denoting the desired projector replacement. For every positive integer $k$, the original gadget hypothesis gives
\[
(B_k)_{00}=(B_k)_{11}=0,\qquad
(B_k)_{01}^2-(B_k)_{10}^2=0.
\]
Each linear expression is a linear functional of $((T|_W)^{\otimes m})^k$. The quadratic expression is a linear functional of $((T|_W)^{\otimes 2m})^k$, obtained by taking two copies of the binary layout with their boundary entries fixed analytically. Apply the annihilating-polynomial argument of the lemma with $m$ and $2m$, respectively. Its nonzero constant coefficient forces the same three equations at $k=0$. Thus $B_0$ is zero or proportional to $\NEq$ or $\SNEq$. The original two classes remain available, so the classes are unchanged. If one of these equations failed for $B_0$, that same polynomial identity would give some $1\le k\le(2m+1)^2$ at which it failed for an actual original binary gadget. This proves preservation for this particular projector interpolation; it is not inferred from Turing equivalence.
\end{proof}

For $q\ge1$, put $E_q=\delta_{0^q1^q}+\delta_{1^q0^q}$. Thus $E_1=\NEq$. From $E_2$, joining one suitable port of $E_q$ to a port of $E_2$ through $\NEq$ gives $E_{q+1}$: the two groups of equal bits each have $q+1$ remaining ports, and each assignment has a unique internal extension. All $E_q$ are therefore available with size linear in $q$.

\begin{lemma}\label{lem:trueq-two-pairs}
Suppose an available balanced quaternary signature, after a local $\SEq$ and a port permutation, is zero outside its middle block $\left(\begin{smallmatrix}p&q\\q&p\end{smallmatrix}\right)$. If $p^2\ne q^2$, then $E_2$ can be adjoined by a projector interpolation that preserves the two binary classes.
\end{lemma}
\begin{proof}
Its transfer is invertible on the middle two coordinates and zero on the other two. Lemma~\ref{lem:trueq-projector} gives the middle block $\NEq$, which is $E_2$ after a port permutation. Corollary~\ref{cor:projector-binary-closure} proves the stated closure.
\end{proof}

The remaining task is to classify all signatures in $\G$ once $E_2$ is available. We use the opposite-pair encoding of~\cite[Section 5]{cai2020eoars}. For non-affine support, the argument below uses a triple of support strings and either an ordinary \#CSP restriction or a six-vertex encoding; it requires no arrow-reversal symmetry.

\begin{lemma}\label{lem:paired-csp-encoding}
Assume $E_2$ is available. For $h\in\G$ of arity $2d$ and a pairing $\pi$ of its ports, define
\[
\phi_{h,\pi}(z_1,\ldots,z_d)
 =h(z_1,1-z_1,\ldots,z_d,1-z_d),
\]
with the ports ordered according to $\pi$. Then $\CSP(\phi_{h,\pi})\le_T\Holant(\NEq\mid\G)$. If every signature $h\in\G$ is supported on opposite bits in one fixed pairing $\pi_h$, put $\Phi=\{\phi_{h,\pi_h}:h\in\G\}$. Then
\[
\Holant(\NEq\mid\G)\equiv_T\CSP(\Phi,\NEq).
\]
\end{lemma}
\begin{proof}
Attach an $E_2$ to each chosen pair of $h$, connecting the two input ports to opposite groups and leaving one port of each group open. The resulting tensor is
\[
\phi_{h,\pi}(z_1,\ldots,z_d)\prod_{j=1}^d\NEq(z_j,z'_j).
\]
Each allowed external assignment has a unique internal extension; this is a restriction gadget, not a formal sum. To simulate a \#CSP variable with $q$ occurrences, use $E_q$, joining each positive/negative port pair to its opposite two groups through $\NEq$. All occurrences then carry the same Boolean value. This realizes the first reduction and, using all signatures in $\G$, the reverse reduction in the displayed equivalence. A logical $\NEq$ constraint is encoded by a port permutation of $E_2$.

For the forward reduction, replace each fixed opposite pair at a vertex by one Boolean variable $z_j$. Each original $\NEq$ edge becomes an equality or disequality between its two logical variables, depending on their port orientations. Identify equality variables and keep disequalities as $\NEq$ constraints. Inconsistent self-constraints give zero. These steps preserve every value. Variables of degree zero contribute two, and the $E_q$ constructions have total size linear in the number of occurrences.
\end{proof}

\begin{theorem}\label{thm:balanced-e2-structure}
Let $\G$ be a finite set of balanced signatures whose nonzero binary gadgets, formed using $\NEq$ edges, have classes $[\NEq],[\SNEq]$, and suppose $E_2$ is available. Then $\Holant(\NEq\mid\G)$ is in FP if $\G\subseteq\mathcal A$ or $\G\subseteq\Prod$, and is \#P-hard otherwise. In the FP alternatives every signature in $\G$ has a fixed pairing on which its two bits are opposite throughout its support.
\end{theorem}
\begin{proof}
First the support of every $h\in\G$ is closed under complementation. For $h$ of arity $2d$ and $x\in\supp(h)$, connect all but one port of $h$ to $E_d$, choosing its two groups according to $\bar x$. Leave one port of each tensor open. Balanced support forces the remaining bit of $h$; the resulting binary has its two off-diagonal entries $h(x),h(\bar x)$, in some order, and zero diagonal. The binary hypothesis gives $h(\bar x)=\pm h(x)\ne0$.

We show that non-affine support gives hardness. Choose $\alpha,\beta,\gamma$ in the support with $\delta=\alpha\oplus\beta\oplus\gamma$ outside it. For each $v\in\mathbb F_2^3$, let $n_v$ count coordinates with column $(\alpha_j,\beta_j,\gamma_j)=v$. Set
\[
D_0=n_{000}-n_{111},\quad D_1=n_{001}-n_{110},\quad
D_2=n_{010}-n_{101},\quad D_3=n_{011}-n_{100}.
\]
The three balanced row sums imply
\[
D_0=D_3=t,\qquad D_1=D_2=-t.
\]
Moreover $|\delta|-d=-2t$.

If $t=0$, pair each column with a complementary column. Then $\alpha,\beta,\gamma,\delta$ all respect that opposite pairing. The compressed signature $\phi_{h,\pi}$ still contains the first three corresponding assignments and omits their XOR. Its support is not affine. It lies in neither $\mathcal A$ nor $\Prod$, so Theorem~\ref{thm:csp-standard} and Lemma~\ref{lem:paired-csp-encoding} give \#P-hardness.

If $t\ne0$, put $m=|t|$. For each of the four complementary column pairs $(v,\bar v)$, attach one $E_{\max(n_v,n_{\bar v})}$ to all its coordinates of $h$. Use one group for the $n_v$ coordinates and the other for the $n_{\bar v}$ coordinates. Exactly $m$ ports remain in a single group. All four output bundles have size $m$, and every allowed assignment is constant on each bundle. Its four bundle values must have weight two, since the whole gadget is balanced. The assignments $\alpha,\beta,\gamma$ and their complements give all six possible weight-two bundle assignments, with nonzero coefficients: each has a unique extension to $h$ and the $E_q$ vertices. Equivalently, for $t>0$ the three bundle strings are $0110,0101,0011$, after the natural group ordering; $t<0$ complements them. Thus this gadget encodes a six-vertex signature with all six weights nonzero.

Replace every vertex of that six-vertex model by this gadget and every edge by $m$ parallel $\NEq$ edges between the corresponding bundles. There is exactly one physical assignment for each logical assignment, and the weights agree. The six-vertex hardness theorem therefore proves that $\Holant(\NEq\mid\G)$ is \#P-hard. The argument is valid for arbitrary $m$; it does not claim that a $4m$-port gadget is an actual quaternary signature.

It remains to consider affine supports. We give the pairing argument explicitly, as in~\cite[Lemma 5.7]{cai2020eoars}. Parametrize a nonempty affine support by $x_j=\ell_j(z)$, with $\ell_j$ affine linear over $\mathbb F_2$. Constant weight gives
\[
\sum_j(-1)^{\ell_j(z)}=0\quad\text{for every }z.
\]
Independence of the Boolean characters implies that for each linear part, equally many coordinates have constant terms zero and one. Pair them. On each pair the bits are opposite on the entire support. This supplies a fixed pairing $\pi_h$ for every $h\in\G$.

Lemma~\ref{lem:paired-csp-encoding} reduces the whole problem to $\CSP(\Phi,\NEq)$ in both directions. Compression by opposite pairs and re-expansion with $\NEq$ preserve $\mathcal A$ and $\Prod$: they substitute complemented variables and multiply by disequality constraints. Since $\NEq$ belongs to both classes, Theorem~\ref{thm:csp-standard} gives exactly the two stated common-class alternatives. Their algorithms are quadratic phase summation over an affine space and propagation of equality/ disequality components with unary weights, respectively.
\end{proof}

\begin{lemma}\label{lem:balanced-decomp-products}
Let $\G$ be a finite set of balanced signatures whose nonzero binary gadgets, formed using $\NEq$ edges, have classes $[\NEq],[\SNEq]$. If every quaternary gadget is decomposable, every gadget is a scalar multiple of a tensor product of $\NEq,\SNEq$. In particular, $\Holant(\NEq\mid\G)$ is in FP.
\end{lemma}
\begin{proof}
The factor analysis in Lemma~\ref{lem:iso-diagonal-factor} leaves only $\mathcal M_4$: a pure extreme power is not balanced, and a product of two nonsingular diagonal factors has nonzero extreme entries. Suppose a minimum-arity gadget outside $\mathcal M$ exists. It is balanced, since every input and every internal $\NEq$ edge is balanced. Its single-pair contractions belong to $\mathcal M$. Lemmas~\ref{lem:iso-high-integration} and~\ref{lem:iso-finite} force it to be a binary product or a signed/permuted RM middle layer at arity eight. The first contradicts its choice. In the second, the actual local normalizations of Section~\ref{ssec:isotropic} give $H_{=}$, and~\eqref{eq:iso-rm-mating} with $a=b=0$ produces a nonzero quaternary supported on two strings. It is not in $\mathcal M_4$, whose nonzero tensors have four support strings. This is a contradiction. Splitting the fixed binary factors reduces closed instances to matrix products around cycles, proving FP.
\end{proof}

\begin{corollary}\label{cor:balanced-c2}
Let $\G$ be a finite set of balanced signatures whose nonzero binary gadgets, formed using $\NEq$ edges, have classes $[\NEq],[\SNEq]$. Then $\Holant(\NEq\mid\G)$ is in FP if $\G\subseteq\mathcal A$ or $\G\subseteq\Prod$, and is \#P-hard otherwise.
\end{corollary}
\begin{proof}
If every quaternary gadget decomposes, Lemma~\ref{lem:balanced-decomp-products} puts the whole signature set in $\mathcal M\subseteq\Prod$. Otherwise normalize an indecomposable quaternary using~\eqref{eq:epsilon}. It has no extreme entries. Three nonzero middle pairs give a hard six-vertex model. One nonzero pair is $E_2$ up to scalar and permutation. With two nonzero pairs, put the middle block in the form $\left(\begin{smallmatrix}a&c\\c&a\end{smallmatrix}\right)$. Indecomposability gives $a^2\ne c^2$, so Lemma~\ref{lem:trueq-two-pairs} adjoins $E_2$ while preserving the two binary classes. Apply Theorem~\ref{thm:balanced-e2-structure}. Since $E_2\in\mathcal A\cap\Prod$, adjoining it does not alter the two common-class conditions on the original signature set.
\end{proof}

\begin{corollary}\label{cor:trueq-one-sided}
Let $\G$ be a finite set of even-arity signatures whose nonzero binary gadgets, formed using $\NEq$ edges, have classes $[\NEq],[\SNEq]$. Suppose all signatures in $\G$ are $\mathrm{HW}^{\ge}$, or all are $\mathrm{HW}^{\le}$. Put $\G_{=}=\{g_{=}:g\in\G\}$. Then $\Holant(\NEq\mid\G)$ is in FP if $\G_{=}\subseteq\mathcal A$ or $\G_{=}\subseteq\Prod$, and is \#P-hard otherwise.
\end{corollary}
\begin{proof}
The boundary weight identity in the proof of Lemma~\ref{lem:iso-one-sided-real} does not require decomposable quaternaries. It gives $q^0=q_{=}$ for every gadget layout and preserves every closed value. Binary tensors are already balanced, so the restricted signature set still has exactly the classes $[\NEq],[\SNEq]$. Apply Corollary~\ref{cor:balanced-c2}. This proves the structure of the restricted signatures before applying their complexity classification.
\end{proof}

The weighted-equality reduction below is the mechanism of~\cite[Lemmas 57--59]{meng2025odd}. We include the equality construction and both complexity directions for the present edge convention.
\begin{lemma}\label{lem:trueq-weighted-equality}
Let $\G$ be a finite signature set. If a weighted equality $e=u\delta_{0^k}+v\delta_{1^k}$, with $uv\ne0$ and fixed $k\ge4$, is realizable from $\G$ using $\NEq$ edges, then $\Holant(\NEq\mid\G)$ is in FP or is \#P-hard.
\end{lemma}
\begin{proof}
A common diagonal transformation $D=\diag(z,z^{-1})$, with $z^{2k}=v/u$, changes $e$ into a nonzero multiple of $=_k$ and satisfies $D^{-T}\NEq D^{-1}=\NEq$. Put $\mathcal H=D\G$. Thus
\[
\Holant(\NEq\mid\G)\equiv_T
\Holant(\NEq\mid\mathcal H,=_k).
\]
We give the equality construction underlying the reduction of Lemma 57 in the published version of~\cite{meng2025odd}. Every $=_{mk}$ can be realized from $=_k$ with $O(m)$ vertices. The assertion is immediate for $m=1$. Given $=_{mk}$, take three new $=_k$ vertices, one as a center and the other two as leaves. Join the center by $\NEq$ edges to one port of $=_{mk}$, one port of the first leaf, and $k-2$ ports of the second leaf. All three outer vertices must have the same bit, opposite to the center's bit. There remain $(mk-1)+(k-1)+2=(m+1)k$ ports, and each of their two constant assignments has a unique internal extension. This proves the induction.

In a \#CSP$_k(\NEq,\mathcal H)$ instance, replace a variable with $mk$ occurrences by this $=_{mk}$ gadget. Connect its ports to the corresponding constraint ports through $\NEq$. Assigning the equality gadget the complement of the CSP variable makes every constraint receive the intended bit. A right-hand $\NEq$ constraint may be suppressed since $\NEq\NEq\NEq=\NEq$. Isolated variables contribute a factor of two. We obtain the value-preserving reduction
\[
\CSP_k(\NEq,\mathcal H)\le_T
\Holant(\NEq\mid\mathcal H,=_k).
\]
Consequently the latter problem is \#P-hard unless $\mathcal H\subseteq\Prod$ or $\mathcal H\subseteq\mathcal A_k^r$ for some $1\le r\le k$, by the \#CSP$_k$ dichotomy.

We verify tractability in both cases. If $\mathcal H\subseteq\Prod$, then $=_k,\NEq\in\Prod$ as well, and propagation of equality and disequality constraints, with unary weights, evaluates the Holant sum in polynomial time. If $\mathcal H\subseteq\mathcal A_k^r$, apply $T_k^{-r}$ to every right-hand signature. All signatures from $\mathcal H$ then belong to $\mathcal A$, while
\[
(T_k^{-r})^{\otimes k}(=_k)
 =\delta_{0^k}+i^{-r}\delta_{1^k}\in\mathcal A,
\qquad
(T_k^r)^T\NEq T_k^r=e^{\pi i r/(2k)}\NEq\in\mathcal A.
\]
The Holant sum is therefore an affine quadratic phase sum and is computable in polynomial time. These arguments apply simultaneously to every signature in $\mathcal H$, completing the dichotomy.
\end{proof}

\begin{lemma}\label{lem:trueq-balanced}
Let $\G$ be a finite set of even-arity signatures whose nonzero binary gadgets, formed using $\NEq$ edges, have classes $[\NEq],[\SNEq]$. If $E_2$ is available, then $\Holant(\NEq\mid\G)$ is in FP or is \#P-hard.
\end{lemma}
\begin{proof}
If every signature in $\G$ is balanced, use Theorem~\ref{thm:balanced-e2-structure}. Otherwise choose $f\in\G$ of arity $n$ and $x\in\operatorname{supp}(f)$ of weight $w\ne n/2$. Put $q=\max(w,n-w)$ and connect all ports of $f$ to $n$ ports of $E_q$, assigning its groups according to $\bar x$. The $k=|2w-n|>0$ remaining ports belong to one group. The resulting actual tensor is
\[
u\delta_{0^k}+v\delta_{1^k},\qquad
\{u,v\}=\{f(x),f(\bar x)\}.
\]
There are no other internal assignments. Since $n$ is even, $k$ is even. The case $k=2$ would be a nonzero diagonal binary, contrary to the proved binary closure. Thus $k\ge4$. If $uv\ne0$, apply Lemma~\ref{lem:trueq-weighted-equality}. If one coefficient is zero, the actual pure extreme power and Lemma~\ref{lem:iso-tail-obstruction} force a common one-sided direction; Corollary~\ref{cor:trueq-one-sided} gives the dichotomy.
\end{proof}

\begin{theorem}[Indecomposable isotropic quaternary signatures]
\label{thm:iso-trueq}\label{thm:5y}
Let $\F$ be a finite even-arity signature set whose nonzero binary gadget classes are $[\Eq],[\SNEq]$. If an indecomposable quaternary gadget exists, then $\Holant(\F)$ is in FP or is \#P-hard.
\end{theorem}
\begin{proof}
Put $\G=K^{-1}\F$ and use $\NEq$ edges. Normalize the middle complementary entries of the chosen gadget using~\eqref{eq:epsilon}:
\[
M(h)=\begin{pmatrix}s&0&0&e\\0&a&c&0\\0&c&a&0\\e&0&0&t\end{pmatrix}.
\]
We first try to obtain $E_2$. Attach $\SEq$ at the first port of a second copy, denoted by $h'$. The actual two-copy gadget is
\begin{equation}\label{eq:trueq-two-copy}
M(h)(\NEq\otimes\NEq)M(h')=
\begin{pmatrix}0&0&0&e^2-st\\0&0&c^2-a^2&0\\
0&a^2-c^2&0&0\\st-e^2&0&0&0\end{pmatrix}.
\end{equation}
A local $\SEq$ equalizes its complementary entries. Permuting the middle pairs of $h$ gives three balanced gadgets with at most two nonzero complementary pairs. With $S=st,A=a^2,C=c^2,E=e^2$, Lemma~\ref{lem:trueq-two-pairs} supplies $E_2$ unless all three equations
\begin{equation}\label{eq:trueq-paired-squares}
(S-E)^2=(C-A)^2,\qquad
(S-A)^2=(C-E)^2,\qquad
(S-C)^2=(E-A)^2
\end{equation}
hold. Its binary closure is preserved by Corollary~\ref{cor:projector-binary-closure}, so Lemma~\ref{lem:trueq-balanced} classifies the whole extension whenever $E_2$ is obtained.

It remains to handle~\eqref{eq:trueq-paired-squares}. Lemma~\ref{lem:paired-squares} allows a middle-pair permutation with $E=S$ and $C=A$. There are only two possibilities, according to whether $S=0$.

If $S=0$, then $e=0$ and at most one extreme entry is nonzero. If $a=0$, the gadget is zero or a pure extreme power, contradicting indecomposability. Otherwise $c=\gamma a$ for $\gamma\in\{1,-1\}$, and the middle part is a scalar multiple of $\NEq\otimes\NEq$ or $\SNEq\otimes\SNEq$ on the pairing $12\mid34$. Local $\SEq$ transformations and scalar normalization make
\[
h=P+u\delta_{b^4},\qquad
P=(\NEq)_{12}(\NEq)_{34},\quad u\ne0,\quad b\in\{0,1\}.
\]
Here $u=0$ would make $h$ decomposable. Put one endpoint of each pair on each side of the transfer matrix. It is $I+uT$ with $T=|bb\rangle\langle(1-b)(1-b)|$ and $T^2=0$, so chains realize $P+ku\delta_{b^4}$ for every positive integer $k$. Lemma~\ref{lem:iso-tail-obstruction} forces one common support direction for $\G$. Corollary~\ref{cor:trueq-one-sided} gives FP exactly when the middle restrictions all belong to $\mathcal A$ or all belong to $\Prod$, and gives \#P-hardness otherwise.

If $S\ne0$, a common diagonal transformation $\diag(z,z^{-1})$ with $z^8=t/s$, followed by scalar normalization of $h$, makes $s=t=1$. This transformation preserves $\NEq$ edges. The two blocks of $M(h)$ are $U_\delta,aU_\gamma$, where
\[
U_\xi=\begin{pmatrix}1&\xi\\\xi&1\end{pmatrix},\qquad
e=\delta\in\{1,-1\},\quad c=\gamma a,\quad\gamma\in\{1,-1\}.
\]
If $a=0$, the sole block has rank one and $h$ decomposes; hence $a\ne0$. Since $U_\xi\NEq=\xi U_\xi$, the two nonzero transfer eigenvalues are $2\delta,2a\gamma$. Lemma~\ref{lem:trueq-projector}, followed by scalar normalization, gives matrix blocks $U_\delta,\delta\gamma U_\gamma$. If $\delta=-1$, attach $\SEq$ at ports 1 and 3. Writing $\eta=\delta\gamma$, the blocks become $U_1,\eta U_\eta$.

For $\eta=1$ this is $[1,0,1,0,1]$. For $\eta=-1$, attaching $\SEq$ at ports 1 and 3 gives $[1,0,-1,0,1]$, and the common diagonal transformation $\diag(1,i)$ changes it to $[1,0,1,0,1]$, preserving $\NEq$ up to scalar. Finally,
\[
K^{\otimes4}[1,0,1,0,1]=2(=_4),\qquad K^{-T}\NEq K^{-1}=I.
\]
Thus the whole transformed problem is an ordinary Holant problem with $=_4$, equivalent to $\CSP_2$ on its entire signature set. Theorem~\ref{thm:csp-even} completes the classification.
\end{proof}

\begin{corollary}[The isotropic dichotomy]\label{thm:isotropic-all}
For every finite even-arity signature set whose nonzero binary gadget classes are $[\Eq],[\SNEq]$, the Holant problem is in FP or is \#P-hard.
\end{corollary}
\begin{proof}
Use Theorem~\ref{thm:iso-complete} when every quaternary gadget decomposes, and Theorem~\ref{thm:iso-trueq} otherwise.
\end{proof}

\section{Conclusion}
\label{sec:conclusion}
We now combine the four branches to prove Theorem~\ref{thm:main}.
\begin{proof}[Proof of the main theorem]
By Lemma~\ref{lem:binary-normal-forms}, an orthogonal transformation puts the binary classes in either normal form. In the normal case, if every quaternary gadget is decomposable, apply Corollary~\ref{thm:normal-decomp-all}; otherwise apply Theorem~\ref{thm:tnm} to an indecomposable quaternary gadget. In the isotropic case, the corresponding alternatives are Theorems~\ref{thm:iso-complete} and~\ref{thm:iso-trueq}. All four cases classify $\Holant(\F)$ and exhaust the possibilities.
\end{proof}

The proof turns a restriction on binary gadgets into structure at every arity. In the decomposable branches, factor propagation and phase sharing supply either binary-product evaluation or a reduction to real signatures. In the indecomposable branches, projector interpolation supplies equalities, and support analysis extends the resulting \#CSP classification to the entire signature set. The needed closure properties are established by tensor-square replacements, polynomial identities, and boundary weight conservation. This gives the stated FP versus \#P-hardness dichotomy under the original $C_2$ hypothesis, including arbitrary tensor factors.

The result settles the cyclic group of order two ($C_2$). The trivial group, larger cyclic groups, and the dihedral and polyhedral groups require further analysis. The four-class extensions arising in the proof connect the $C_2$ problem with the Klein-group setting and suggest methods for studying larger finite binary groups.

\section*{AI methodologies and Acknowledgments}
\paragraph{AI methodologies}We used an AI assistant to assist with checking the correctness of the whole proof, including Sections~4 and~5, and in particular with the proof concerning decomposable quaternary signatures in isotropic cases. The tool helped correct the structure of signatures when all quaternary signatures are decomposable and binary signatures are isotropic in Section~\ref{ssec:isotropic}, and helped check and simplify the proofs in Section~\ref{ssec:normal5} (the nonzero-extreme-entry branch) and Section~\ref{ssec:trueq-isotropic} (the balanced-equality reduction and the residual cases). AI was also used to improve the writing throughout the paper. The tool materially affected mainly Sections~\ref{ssec:isotropic} and ~\ref{ssec:trueq-isotropic}. The authors verified the correctness and originality of all content, including references.

\begingroup\emergencystretch=2em
\paragraph{Acknowledgments}The author is supported by the National Key R\&D Program of China (2023YFA1009500), NSFC 62532014 and NSFC 62272448. We are deeply grateful to our advisor, Mingji Xia, for his guidance and support throughout this work. He provided valuable insights on Pauli bases and real-numberization methods, which greatly facilitated the proof of the decomposable cases. Despite his contributions, he generously declined to be a coauthor. We would like to thank Jiayi Zheng for insightful discussions that helped us complete the overall framework, especially for the isotropic cases. We also appreciate Yuliang Nie and Yu Chen for their help in algebra.
\par\endgroup

\bibliography{tbib}

\end{document}